# Lower Bounds for the Complexity of Monadic Second-Order Logic

Stephan Kreutzer
University of Oxford
kreutzer@comlab.ox.ac.uk

Siamak Tazari
Humboldt Universität zu Berlin
tazari@informatik.hu-berlin.de

## Abstract

*Courcelle's famous theorem from 1990 states that any property of graphs definable in monadic second-order logic ($MSO_2$) can be decided in linear time on any class of graphs of bounded treewidth, or in other words, $MSO_2$ is fixed-parameter tractable in linear time on any such class of graphs. From a logical perspective, Courcelle's theorem establishes a sufficient condition, or an upper bound, for tractability of $MSO_2$-model checking.*

*Whereas such upper bounds on the complexity of logics have received significant attention in the literature, almost nothing is known about corresponding lower bounds. In this paper we establish a strong lower bound for the complexity of monadic second-order logic. In particular, we show that if $\mathcal{C}$ is any class of graphs which is closed under taking subgraphs and whose treewidth is not bounded by a polylogarithmic function (in fact, $\log^c n$ for some small $c$ suffices) then $MSO_2$-model checking is intractable on $\mathcal{C}$ (under a suitable assumption from complexity theory).*

## 1 Introduction

In 1990, Courcelle proved a fundamental result stating that every property of graphs definable in *monadic second-order logic with edge set quantification* ($MSO_2$), the extension of first-order logic by quantification over sets of vertices and edges, can be decided in linear time on any class $\mathcal{C}$ of graphs of bounded treewidth. This theorem has important consequences both in logic and in algorithm theory. In the theory of efficient algorithms on graphs, it can often be used as a simple way of establishing that a property can be solved in linear time on graph classes of bounded treewidth. Besides being of interest for specific algorithmic problems, results such as Courcelle's and similar *algorithmic meta-theorems* lead to a better understanding how far certain algorithmic techniques, such as dynamic programming on bounded treewidth graphs, range; and also establish general upper bounds for the parameterised complexity of a wide range of problems. See [Gro07, Kre09a] for recent surveys on algorithmic meta-theorems.

From a logical perspective, Courcelle's theorem establishes a sufficient condition for tractability of $MSO_2$ formula evaluation on classes of graphs or structures: whatever the class $\mathcal{C}$ may look like, if it has bounded treewidth, then $MSO_2$-model checking is tractable on $\mathcal{C}$. An obvious question is how tight the theorem actually is, i.e. whether it can be extended to classes of unbounded treewidth and if so, how large the treewidth of graphs in the class can be in general. Given the considerable interest in Courcelle's theorem, and the far-reaching consequences that extensions of this result to interesting classes of graphs of unbounded treewidth would have, it is surprising that not much is known about such limits for $MSO_2$-model checking. To fully understand the (parameterised) complexity of monadic second-order logic with respect to particular classes of graphs, we need to understand necessary conditions for tractability as much as sufficient conditions; but for some reason necessary conditions have so far not been studied in much depth.

In order to formally state and further discuss our results, also in relation to previous work, we need the following notion; it basically states that a class of graphs of unbounded treewidth actually contains sufficiently many graphs *witnessing* the large treewidth of the class and that these witnesses can be constructed efficiently.

**Definition 1.1.** The treewidth of a class $\mathcal{C}$ of graphs is *strongly unbounded* by a function $f : \mathbb{N} \to \mathbb{N}$ if there is $\varepsilon < 1$ and a polynomial $p(x)$ such that for all $n \in \mathbb{N}$ there is a graph $G_n \in \mathcal{C}$ with

1. the treewidth of $G_n$ is between $n$ and $p(n)$ and is not bounded by $f(|G_n|)$ and
2. given $n$, $G_n$ can be constructed in time $2^{n^\varepsilon}$.

The degree of the polynomial $p$ is called the *gap-degree* of $\mathcal{C}$ (with respect to $f$). The treewidth of $\mathcal{C}$ is *strongly unbounded polylogarithmically* if it is strongly unbounded by $\log^c n$, for all $c \geq 1$.

A first lower bound for the complexity of monadic second-order logic appeared in [Kre09b] and has been extended in [KT10]. In these papers, it was shown that $MSO_2$-model-checking is not fixed-parameter tractable on any class of graphs where

a) the treewidth is strongly unbounded by $\log^{28} n$ and
b) which are *closed under re-colourings* for a fixed set $\Gamma$ of colours, i.e. if $G \in \mathcal{C}$ and $G'$ is obtained from $G$ by colouring some vertices or edges by colours from $\Gamma$, then $G' \in \mathcal{C}$.

These papers establish powerful logical and algorithmic tools for proving such intractability results and we will resort to some of these tools below. However, closure under colourings is a very strong condition as it allows to "mark" bad substructures in a graph. In this work we aim for an even stronger intractability result for $MSO_2$:

**Theorem 1.2.** *Let $\mathcal{C}$ be a class of graphs closed under subgraphs, i.e. $G \in \mathcal{C}$ and $H \subseteq G$ implies $H \in \mathcal{C}$.*

1. *If the treewidth of $\mathcal{C}$ is strongly unbounded by $\log^{28\gamma} n$, where $\gamma > 1$ is larger than the gap-degree of $\mathcal{C}$, then $MC(MSO_2, \mathcal{C})$ is not in XP, and hence not fixed-parameter tractable, unless SAT can be solved in subexponential time.*

2. *If the treewidth of $\mathcal{C}$ is strongly unbounded polylogarithmically then $MC(MSO_2, \mathcal{C})$ is not in XP unless all problems in the polynomial-time hierarchy can be solved in subexponential time.*

Recall that $MC(MSO_2, \mathcal{C})$ refers to the parameterised model-checking problem for $MSO_2$. We will give a justification for the two conditions in Definition 1.1 below, once we have discussed some more related work. To give an example, the theorem implies that the class $\mathcal{C}$ of all (or all planar, bipartite, etc.) graphs $G$ of treewidth $tw(G) \leq \log^{29} |G|$ does not have fixed-parameter tractable $MSO_2$ model-checking unless SAT can be solved in subexponential time.

## 1.1 Related Work

Theorem 1.2 complements the intractability result of [Kre09b, KT10] in that it refers to classes of graphs closed under subgraphs and does not require any colours, a much more natural condition.

In [Gro07, Conjecture 8.3], Grohe conjectures [1] the following.

**Conjecture 1.3** (Grohe [Gro07]). *Let $\mathcal{C}$ be a class of graphs that is closed under taking subgraphs. Suppose that the treewidth of $\mathcal{C}$ is not polylogarithmically bounded, that is, there is no constant $c$ such that $tw(G) \leq \log^c |G|$ for every $G \in \mathcal{C}$. Then the model-checking problem of $MSO_2$ is not fixed parameter tractable on $\mathcal{C}$.*

---
[1]The original conjecture is formulated in terms of branchwidth but this is equivalent.

Clearly, with current technology there is no hope to prove any such conjecture without relating it to assumptions in complexity theory (as the conjecture implies $\mathsf{P} \neq \mathsf{PSPACE}$). In this sense, our result only proves Grohe's conjecture modulo complexity theoretical assumptions and the additional conditions on strongly unboundedness necessitated by this. On the other hand, our result is stronger than the conjecture in that we only require a fixed log-power rather than polylog.

In [MM03], Makowsky and Mariño study similar questions in relation to classes of graphs closed under topological minors. They show that any such class must have bounded treewidth for $\mathrm{MSO}_2$ model-checking to be in FPT. Closure under topological minors is a much stronger condition simplifying the proof significantly. However, in the same paper, the authors give examples for classes of graphs of unbounded clique-width but with tractable $\mathrm{MSO}_1$ model-checking. These examples can be adapted to examples of classes of graphs which are closed under subgraphs, whose treewidth is only bounded logarithmically (but which almost have logarithmic treewidth) and on which $\mathrm{MSO}_2$ model-checking is tractable. This shows that in full generality, our results can not be strengthened much beyond the $\log^{28\gamma} n$ bound postulated in Theorem 1.2.

## 1.2 On Strongly Unbounded Treewidth

Let us give some justification for the two conditions in Definition 1.1. The first condition is a consequence of the fact that we prove our main result by reducing an NP-hard problem to $\mathrm{MC}(\mathrm{MSO}_2, \mathcal{C})$. Without this condition there could simply be too few graphs of high treewidth in $\mathcal{C}$ to define a reduction. To give an example, fix a constant $c$ and let $H_n$ be the graph constructed from the $n \times n$-grid by replacing every edge by a path on $\frac{1}{m} \cdot 2^{\sqrt[c]{n}}$ vertices, where $m = n^2$. The resulting graph has $\mathcal{O}(2^{\sqrt[c]{n}})$ vertices and treewidth $n$. Now let $\mathcal{C}' := \{H_n : n = 2^{2^i}, i > 0\}$ and let $\mathcal{C}$ be the subgraph closure of $\mathcal{C}'$. If $c > 29$, then the treewidth of $\mathcal{C}$ is unbounded by $\log^{29} n$ but not strongly unbounded by this function, while being closed under taking subgraphs. To see this, take a graph $H_n \in \mathcal{C}'$, for some $n = 2^{2^i}$, $i > 2$. Any subgraph $H \subseteq H_n$ is either acyclic, and therefore has treewidth 1, or it contains a path of length $\frac{1}{m} \cdot 2^{\sqrt[c]{n}}$. Thus, $H_n$ does not contain any subgraph $H \subseteq H_n$ of treewidth $2^i \leq \mathrm{tw}(H) \leq p(2^i)$ such that $\mathrm{tw}(H) > \log^c |H|$, for any fixed polynomial $p$. It follows that if we wanted to use $\mathcal{C}$ for a reduction as outlined below, there wouldn't be enough graphs of large treewidth to reduce to: given an instance of SAT of length $2^i$ for an $i$ that is not close to a power of 2, we would have no chance in identifying a graph in $\mathcal{C}$ to perform a reduction in polynomial time. Therefore, as long as we have to rely on reductions to prove results as in this work, a condition similar to Condition 1 seems necessary.

The second condition is necessary to prevent artificial cases where constructing a graph in the class $\mathcal{C}$ is already so expensive that any reduction would take too much time.

## 1.3 Overview of the Proof

Let us briefly sketch the main ideas of the proof, the basic framework of which is adapted from [Kre09b]. Let $\mathcal{C}$ be a class of graphs with treewidth strongly unbounded by $\log^c n$, for some suitable $c$.

We aim at reducing the propositional satisfiability problem SAT to $\mathrm{MC}(\mathrm{MSO}_2, \mathcal{C})$. Towards this aim we will first construct an $\mathrm{MSO}_2$-formula $\varphi$, depending only on a Turing machine deciding SAT, and then, given a SAT-instance $w$, construct a graph $G_w \in \mathcal{C}$ such that $G_w \models \varphi$ if and only if $w$ is satisfiable. The idea is to encode the instance $w$ in the graph $G_w$ so that (i) the instance can be decoded by the $\mathrm{MSO}_2$-formula $\varphi$ and (ii) the graph $G_w$ contains enough structure so that the formula $\varphi$ can simulate the run of a Turing machine deciding SAT on input $w$.

Similar ideas in connection with treewidth have been employed in the past and the usual approach is to use the the Excluded Grid Theorem of Robertson, Seymour, and Thomas [RST94] that there is a function $f : \mathbb{N} \to \mathbb{N}$ such that every graph of treewidth $f(k)$ contains a $k \times k$-grid as a minor. Such a grid provides enough structure to simulate runs of Turing machines in MSO$_2$ and encoding the SAT instance $w$ in a grid can easily be done by deleting certain edges (see Section 6).

However, the best known bound for the function $f$ known to date is exponential and as we are dealing with graphs of treewidth only logarithmic in the number of vertices, the grids we are guaranteed to find in this way are essentially only of order $\log \log |G_w|$ which is much too small for any reduction to work.

Instead of using grids, therefore, we will use a new structural characterisation of treewidth developed by Reed and Wood [RW08] and made algorithmic in [KT10] which replaces grids by *grid-like minors*. It was shown in [RW08] that any graph contains a grid-like minor of order polynomial in its treewidth and in [KT10] it was shown that these are computable in polynomial time. The main problem with grid-like minors is that a) they can resemble cliques rather than grids and b) they do not occur as minors of the graph itself but only of the intersection graph of sets of pairwise disjoint paths (see Section 2 for details). As indicated above, we would like to encode a SAT-instance $w$ in a grid by deleting certain edges. But as grid-like minors only occur as minors of intersection graphs, deleting an edge in a graph $G$ has no predictable implication for the grid-like minor which makes encoding SAT-instances using grid-like minors extremely difficult.

Therefore, instead of encoding SAT-instances in grid-like minors directly, we will impose a labelling of the grid-like minor externally. For this, given a SAT-instance $w := w_1 \ldots w_l$ and a graph $G$ of sufficiently high treewidth, we construct a tree $T \subseteq G$ which has a special structure so that there is an MSO$_2$-formula defining a linear order on trees of this structure. Furthermore, this particular structure of the tree allows us to encode the letters $w_i$ in subtrees of $T$ containing some of the leaves (we will call these *single crosses* (encoding 0) and *double crosses* (encoding 1)). Hence, the order imposed on $T$ together with the ability to encode letters allows us to encode the SAT-instance $w$ in $T$. We will then show that $G$ also contains a grid-like minor which is attached to the tree $T$ so that the word encoded in $T$ can be transferred to a unique labelling of the grid-like minor. Hence, we will use this external tree to encode the SAT-instance and the grid-like minor as the structure we need to simulate the run of a Turing-machine on the encoded input. The tree $T$ together with the grid-like minor attached to it is called a *labelled tree-ordered web* and is illustrated in Figure 2.

Finally, as we assume that the class $\mathcal{C}$ of graphs we work in is closed under subgraphs, this labelled tree-ordered web occurs as a graph in $\mathcal{C}$. Hence, if evaluating the MSO$_2$-formula which decodes the encoded SAT-instance and simulates the run of a Turing machine on it was fixed-parameter tractable, we could solve SAT in sub exponential time.

**Organisation.** The paper is organised as follows. In Section 2 we recall notation and concepts from graph theory. We recall monadic second-order logic in Section 3. In Section 4 we define the labelled tree-ordered webs discussed above and show that every graph of sufficient treewidth contains such a structure. This is the main algorithmic result of the paper.

We continue in Section 5 by defining the various parts and the order of a labelled tree-ordered web in MSO$_2$. In Section 6, we review the notion of MSO$_2$ − MSO$_2$-interpretations and the hardness of MSO$_2$ on coloured walls before showing its hardness on uncoloured walls; afterwards, we show how to define an interpretation of a labelled tree-ordered web in a coloured wall. Finally, we present the proof of Theorem 1.2 in Section 7 and conclude in Section 8.

## 2 Preliminaries

We use standard notation from graph theory and refer to [Die05] for details. In particular, all graphs in this paper are simple and undirected. We write $V(G)$ and $E(G)$ for the set of vertices and edges of a graph $G$. W.l.o.g. we assume $V(G) \cap E(G) = \varnothing$. A *path* $P \subseteq G$ in a graph $G$ is a connected acyclic subgraph in which every vertex has degree at most 2.

Treewidth is a global connectivity measure of graphs that was introduced by Robertson and Seymour in their graph minor series. Essentially, it associates to each graph $G$ a number $\operatorname{tw}(G) \in \mathbb{N}$ measuring how similar a graph is to being a tree. We will not need the precise definition in this paper and therefore refer the reader to [Die05] for a definition of treewidth.

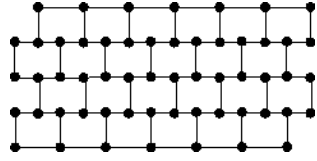

**Figure 1. Elementary** $4 \times 6$**-wall.**

**Definition 2.1.** Let $f : \mathbb{N} \to \mathbb{N}$ be a function and $\mathcal{C}$ be a class of graphs. The treewidth of $\mathcal{C}$ is *bounded by* $f$, if $\operatorname{tw}(G) \leq f(|G|)$ for all $G \in \mathcal{C}$. $\mathcal{C}$ has *bounded treewidth* if its treewidth is bounded by a constant.

Many natural classes of graphs, for instance series-parallel graphs, are found to have bounded treewidth.

**Definition 2.2.** A *subdivision* of a graph $H$ is a graph $H'$ obtained from $H$ by iteratively replacing some edges by paths of length 2. The original vertices of $H$ in $H'$ are called the *nails* of $H$ in $H'$. If a graph $G$ contains a subdivision of $H$, we call $H$ a *topological minor* of $G$.

**Definition 2.3.** Let $n, m > 0$ be integers. An *elementary* $n \times m$-*wall* is a graph with vertex set $V :=$ $\{(1, 2j-1) : 1 \leq j \leq m+1\} \cup \{(i,j) : 1 < i \leq n, 1 \leq j \leq 2m+2\} \cup$ $\{(n+1, 2j-t) : 1 \leq j \leq m+1 \text{ and } t := (n \bmod 2)\}$ and edge set

$$\begin{aligned}
E := {} & \{(1, 2j-1), (1, 2(j+1)-1) : 1 \leq j \leq m\} \quad \text{// horizontal edges in row 1} \\
& \cup \{(n+1, 2j-t), (n+1, 2(j+1)-t) : 1 \leq j \leq m \text{ and } t := (n \bmod 2)\} \\
& \quad \text{// horizontal edges in row } n+1 \\
& \cup \{(i,j), (i, j+1) : 2 \leq i \leq n, 1 \leq j < 2m+2\} \quad \text{// further horizontal edges} \\
& \cup \{(i,j), (i+1, j) : 1 \leq i \leq n, 1 \leq j \leq 2m+2, i \text{ and } j \text{ even }\} \quad \text{// vertical edges} \\
& \cup \{(i,j), (i+1, j) : 1 \leq i \leq n, 1 \leq j \leq 2m+2, i \text{ and } j \text{ odd }\}. \quad \text{// vertical edges}
\end{aligned}$$

An $n \times m$-*wall* is a subdivision of an elementary $n \times m$-wall. The *nails* of a wall are the nails of the subdivision of the elementary wall it is obtained from. An elementary wall is a graph as illustrated in Figure 1.

We will always think of the vertices of a wall as being numbered in a way that $(1,1)$ is the vertex in the "bottom-left corner". The "bottom-row" of an $n \times m$-matrix is then the row 1.

Let $\mathcal{P}$ and $\mathcal{Q}$ each be a set of disjoint paths of a graph $G$. We denote by $\mathcal{I}(\mathcal{P}, \mathcal{Q})$ the *intersection graph* of $\mathcal{P}$ and $\mathcal{Q}$ defined as the bipartite graph with vertex set $\mathcal{P} \cup \mathcal{Q}$ and an edge between two vertices if and only if the corresponding paths intersect. The following definition is adapted from Reed and Wood's [RW08] definition of a *grid-like minor*:

**Definition 2.4.** Let $\mathcal{P}$ and $\mathcal{Q}$ be each a set of disjoint paths in a graph $G$. $(\mathcal{P}, \mathcal{Q})$ is called a *topological grid-like minor of order* $\ell$ in $G$ if $\mathcal{I}(\mathcal{P}, \mathcal{Q})$ contains a subdivision of the complete graph $K_\ell$. The *nails* of $(\mathcal{P}, \mathcal{Q})$ are the paths corresponding to the nails of the subdivision of $K_\ell$ in $\mathcal{I}(\mathcal{P}, \mathcal{Q})$.

## 3 Monadic Second-Order Logic

In this section we briefly recall the definition of monadic second-order logic. As we are mainly interested in graphs we only introduce MSO$_2$ on graphs.

The class of formulas of *monadic second-order logic with edge set quantification*, denoted MSO$_2$, is defined as the extension of first-order logic by quantification over sets of edges and sets of vertices. However, for the purpose of this paper it is more convenient to define it formally as monadic second-order logic on *incidence structures*.

**Signatures and Structures.** A signature $\sigma$ is a finite set of relation symbols $R$ each of arity $ar(R)$. A $\sigma$-structure $A$ consists of a *universe* $U(A)$ and for each $R \in \sigma$ an $ar(R)$-ary relation $R(A) \subseteq (U(A))^{ar(R)}$.

**Incidence Structures.** The signature $\sigma_{graph}$ of incidence structures is defined as $\sigma_{graph} := \{V, E, \in\}$, where $V, E$ are unary and $\in$ is a binary relation symbol. We will always use $\in$ in infix notation and write $v \in^A e$ instead of $(v, e) \in\in (A)$. With any graph $G$ we associate a $\sigma_{graph}$-structure $A := \mathcal{G}(G)$, its *incidence structure*, with universe $U(A) := V(G) \dot\cup E(G)$ and $V(A) := V(G)$, $E(A) := E(G)$ and $v \in^A e$ if $v \in V(G), e \in E(G)$ and $v$ and $e$ are incident in $G$. We will not usually distinguish between a graph $G$ and its incidence structure.

**Monadic Second-Order Logic (MSO$_2$).** MSO$_2$ is the extension of first-order logic by quantification over sets of elements (which can be vertices or edges). That is, in addition to first-order variables, which we will denote by small letters $x, y, ...$, there are variables $X, Y, ...$ ranging over sets of elements. Formulas of MSO$_2[\sigma]$ are then build up inductively by the rules for first-order logic FO$[\sigma]$ with the following additional rules: if $X$ is a second-order variable and $\varphi \in \text{MSO}_2[\sigma \dot\cup \{X\}]$, then $\exists X \varphi \in \text{MSO}_2[\sigma]$ and $\forall X \varphi \in \text{MSO}_2[\sigma]$ with the obvious semantics where, e.g., a formula $\exists X \varphi$ is true in a $\sigma$-structure $G$ if there is a subset $X' \subseteq U(G)$ such that $\varphi$ is true in $G$ if the variable $X$ is interpreted by $X'$. We write $G \models \varphi$ to indicate that $\varphi$ is true in $G$.

If $\varphi(x)$ is a formula with a free first-order variable $x$ and $G$ is a structure, we write $\varphi(G)$ for the set $\{v \in U(G) : G \models \varphi[v]\}$. See [Lib04] for more on MSO$_2$.

We state some conventions and give some examples of MSO$_2$-formulas that we are going to need later:

- for terms $s, t$, we write $s \neq t$ for $\neg(s = t)$;
- we write $\top$ for the formula $\forall x\, x = x$ and $\bot$ for the formula $\forall x\, x \neq x$;
- for formulas $\varphi, \psi$, we write $(\varphi \to \psi)$ for $(\neg \varphi \vee \psi)$;
- for formulas $\varphi, \psi$, we write $(\varphi \leftrightarrow \psi)$ for $(\varphi \to \psi) \wedge (\psi \to \varphi)$;
- $\neg$ is evaluated first, $\vee$ and $\wedge$ are evaluated next, followed by the quantifiers $\exists, \forall$, and $\to, \leftrightarrow$ are evaluated last; we often omit the outermost parentheses of a formula;
- $\bigvee_{i=1}^{k} \varphi(x_i)$ denotes $\varphi(x_i) \vee \cdots \vee \varphi(x_k)$; we freely use such variations with $\bigvee$ and $\bigwedge$;
- $\exists^{\leq k} x \varphi(x)$ denotes $\exists x_1 \ldots \exists x_k \neg \exists y (\varphi(y) \wedge \bigwedge_{i=1}^{k} y \neq x_i)$ intending to mean that there exist at most $k$ elements in the structure fulfilling $\varphi(x)$;
- $\exists^{\geq k} x \varphi(x)$ to mean that there exist at least $k$ elements in the structure fulfilling $\varphi(x)$, i.e. $\neg \exists^{\leq k-1} x\, \varphi(x)$;
- $\exists^{=k} x \varphi(x)$ to mean that there exist exactly $k$ elements in the structure fulfilling $\varphi(x)$, i.e. $\exists^{\leq k} x\, \varphi(x) \wedge \exists^{\geq k} x\, \varphi(x)$;
- for a second-order variable $P$ denoting a set of edges, we write $x \in V(P)$ for the formula $\exists e(e \in P \wedge v \in e)$;
- we can define that $X$ consists of connected components of $P$ by

$$\textit{components}(X, P) := X \subseteq P \subseteq E \wedge \forall e \in P (e \cap V(X) \neq \varnothing \to e \in X);$$

- $\textit{conn}(P) := \forall X \neq \varnothing (\textit{components}(X, P) \to X = P)$ states that $P$ is connected;
- $\textit{deg}^{\leq k}(v, P) := \exists^{\leq k} e \in P(v \in e)$ states that the degree of a vertex $v$ is at most $k$ in $P$; similarly, $\textit{deg}^{\geq k}(v, P)$ and $\textit{deg}^{=k}(v, P)$ can be defined to state that the degree of a vertex $v$ is at least or exactly $k$ in $P$, respectively;
- $\textit{ac}(P) := \forall X \subseteq P \forall e \in X \forall u, v \in e \big(\textit{conn}(X) \wedge u \neq v \wedge \textit{deg}^{\geq 2}(u, X) \wedge \textit{deg}^{\geq 2}(v, X) \to \forall Y(Y = X - e \to \neg \textit{conn}(Y))\big)$ states that $P$ is acyclic; and

- we can define that $P$ is path by

$$path(P) := conn(P) \land ac(P) \land \forall v\, deg^{\leq 2}(v, P)\,.$$

**Model Checking.** The *model checking problem* $\text{MC}(\text{MSO}_2)$ for $\text{MSO}_2$ is defined as the problem, given a structure $G$ and a formula $\varphi \in \text{MSO}_2$, to decide if $G \models \varphi$. In [Var82], Vardi proved that $\text{MC}(\text{MSO}_2)$ is PSPACE-complete. However the hardness result crucially uses the fact that the formula is part of the input (and in fact holds on a fixed two-element structure), whereas we are primarily interested in the complexity of checking a fixed formula expressing a graph property in a given input graph. We therefore study model-checking problems in the framework of *parameterised complexity* (see [FG06] for background on parameterised complexity).

**Definition 3.1.** Let $\mathcal{C}$ be a class of $\sigma$-structures. The *parameterised model-checking problem* $\text{MC}(\text{MSO}_2, \mathcal{C})$ for $\text{MSO}_2$ on $\mathcal{C}$ is defined as the problem to decide, given $G \in \mathcal{C}$ and $\varphi \in \text{MSO}_2[\sigma]$, if $G \models \varphi$. The *parameter* is $|\varphi|$.

$\text{MC}(\text{MSO}_2, \mathcal{C})$ is *fixed-parameter tractable* (fpt), if for all $G \in \mathcal{C}$ and $\varphi \in \text{MSO}_2[\sigma]$, $G \models \varphi$ can be decided in time $f(|\varphi|) \cdot |G|^k$, for some computable function $f$ and $k \in \mathbb{N}$. The problem is in the class XP, if it can be decided in time $|G|^{f(|\varphi|)}$.

An important aspect of parameterised complexity is that it is invariant under syntactic variations of the logic, i.e. if $\mathcal{L}$ and $\mathcal{L}'$ are equivalent in the sense that formulas of one logic can effectively be translated into equivalent formulas of the other logic, then $\mathcal{L}$ is fpt on a class $\mathcal{C}$ if, and only if, $\mathcal{L}'$ is fpt on $\mathcal{C}$. The corresponding statement is false for classical complexity.

As, for instance, the NP-complete problem 3-Colourability is definable in $\text{MSO}_2$, $\text{MC}(\text{MSO}_2, \text{GRAPHS})$, the model-checking problem for $\text{MSO}_2$ on the class of all graphs, is not fixed-parameter tractable unless $P = \text{NP}$. However, Courcelle proved that if we restrict the class of admissible input graphs, then we can obtain much better results.

**Theorem 3.2** ([Cou90]). $\text{MC}(\text{MSO}_2, \mathcal{C})$ *is fixed-parameter tractable on any class $\mathcal{C}$ of graphs of treewidth bounded by a constant.*

## 4 Labelled Tree-Ordered Webs

The goal of this section is to prove the main algorithmic aspects of this paper. As indicated in the introduction, we aim at encoding instances $w$ of an NP-hard problem $P$, i.e. $w$ is a word over the alphabet $\{0, 1\}$, in graphs of large enough tree-width. The core algorithmic problem is to identify a structure so that there is a polynomial $p(n)$ such that given a word $w$ of length $m$ and a graph of tree-width at least $p(m)$, $G$ contains such a structure encoding $w$ as a subgraph. The structure we are after, which we call *labelled tree-ordered web*, is indicated in Figure 2. It consists of a tree $T$, which will have a special structure so that there is an $\text{MSO}_2$-formula defining a linear order on the vertices of such trees. Some vertices of the tree will be adjacent to what we call *single crosses* and *double crosses* – indicated by the black vertices in Figure 2 – which encode the letters 0 and 1. Furthermore, there is a part which consists of two sets $\mathcal{P}, \mathcal{Q}$ of disjoint paths so that their intersection graph $\mathcal{I}(\mathcal{P}, \mathcal{Q})$ contains a wall in which the paths constituting the bottom-row are adjacent to vertices in $T$. The idea is that the crosses encode the word $w$. In between any two crosses (with respect to the definable tree-ordering) there is a vertex connected to an element on the bottom-row of the wall and in this way we induce a labelling of the bottom-row of this "grid-like minor" by the word $w$. All this will be decodable in $\text{MSO}_2$ and once this is done we can guess in $\text{MSO}_2$ an accepting run of a Turing-machine deciding the problem $P$ on input $w$.

Starting with a certain subgraph provided in [KT10] that contains a grid-like minor of large order, we incrementally modify it in the subsections below, introducing more and more structure into it, until we obtain the desired subgraph.

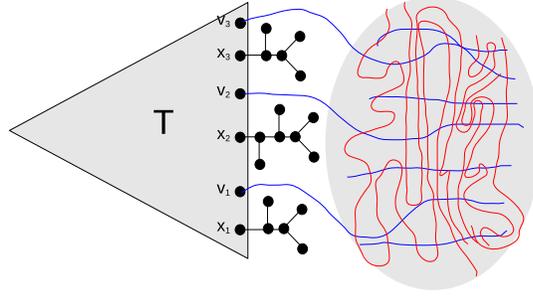

**Figure 2. A labelled tree-ordered web encoding** `010`.

### 4.1 $k$-Webs

We start by reviewing the structure of a *$k$-web of order $h$.* as introduced in [KT10], and its algorithmic properties.

**Definition 4.1.** A *$k$-web of order $h$* in a graph $G$ is a collection $(T, (T_i)_{1 \le i \le h}, (A_i)_{1 \le i \le h}, B)$ of subgraphs of $G$ such that

1. $T$ is a subcubic tree and $V(B \cap T) = \bigcup_{1 \le i \le h} V(A_i)$;
2. $T_1, \ldots, T_h$ are disjoint subtrees of $T$ and for $1 \le i \le h$, $A_i \subseteq T_i$ is flat in $T$;
3. for all $1 \le i < j \le h$ there is a set $\mathcal{P}_{i,j}$ of $k$ disjoint paths in $B$ connecting $A_i$ and $A_j$;

A *model* of $H$ in $G$ is a map that assigns to every vertex of $H$, a connected subgraph of $G$, such that the images of the vertices of $H$ are all disjoint in $G$ and there is an edge between them if there is an edge between the corresponding vertices in $H$. A graph $H$ is a *minor* of $G$ if and only if $G$ contains a model of $H$. A topological minor of $G$ is also a minor of $G$ but the reverse is not true in general. However, if $H$ has maximum degree 3, then $H$ is a minor of $G$ if and only if it is a topological minor of $G$.

**Theorem 4.2.** *(adapted from Theorem 4.2 in [KT10]) Let $G$ be a graph of tree-width at least $c\ell^7$, for a constant $c$. Then $G$ contains either a model of $K_\ell$ or, for some $k$, a $k$-web of order $4$ that contains a topological grid-like minor of order $\ell$. Furthermore, either outcome can be computed in polynomial time.*

### 4.2 Tree-Webs

The notion of a *tree-web*, defined below, is central to this part of our work; in the subsequent sections, we will gradually refine this notion until we finally obtain the structure that we need.

**Definition 4.3.** A *tree-web of order $\ell$* is a tuple $\mathcal{W} = (G, T, r, A, \mathcal{P}, \mathcal{Q})$, so that

1. $T$ is a subcubic tree rooted at $r$,
2. $(\mathcal{P}, \mathcal{Q})$ is a topological grid-like minor of order $\ell^2$ whose nails are paths from $\mathcal{P}$,
3. $G$ is a graph of maximum degree 4 with $T \cup \bigcup \mathcal{P} \cup \bigcup \mathcal{Q} \subseteq G$,
4. $T$ only intersects with nails of $(\mathcal{P}, \mathcal{Q})$,
5. the paths from $\mathcal{P}$ that are nails are either disjoint from $T$ or intersect $T$ in exactly one endpoint, and
6. $A = V(T) \cap V(\bigcup \mathcal{P})$ is flat in $T$.

The vertices of $A$ are called the *good* vertices of $\mathcal{W}$. The paths in $\mathcal{P}$ that start at a vertex in $A$ are called *good* paths.

In case $G = T \cup \bigcup \mathcal{P} \cup \bigcup \mathcal{Q}$ and *all* the nails in $\mathcal{P}$ are good, and hence intersect $T$, i.e. $|A| = \ell^2$, we call the structure a *full tree-web*. A *subtree-web* of a tree-web $\mathcal{W} = (G, T, r, A, \mathcal{P}, \mathcal{Q})$ is a tree-web $\mathcal{W}' = (G', T', r', A', \mathcal{P}', \mathcal{Q}')$ with $G' \subseteq G$. In this case, we write $\mathcal{W}' \subseteq \mathcal{W}$. A *full subtree* of a rooted tree $(T, r)$ is the connected component of $T - e$ not containing $r$, for some $e \in E(T)$.

**Definition 4.4.** A tree-web $\mathcal{W} = (G, T, r, A, \mathcal{P}, \mathcal{Q})$ is *nice* if

1. $T \cup \bigcup \mathcal{P} \cup \bigcup \mathcal{Q}$ has no vertex of degree 1 except maybe $r$,
2. if $P = v_0, \ldots, v_k$ is a good path with $v_0 \in A$, $v_1$ does not lie on any other path,
3. every full subtree of $T$ with at least 2 vertices contains at least 2 good vertices.
4. every leaf of $T$ is good, and
5. the neighbour of every leaf of $T$ in $T$ is good.

Note that the last two conditions are implied by the third. The proof of Lemma 4.6 below is based on the combinatorial Lemma 4.5.

**Lemma 4.5.** *Let $G := \mathbf{K}_k$ be a clique on $k$ vertices and assume at most $k$ edges of $G$ are coloured red and the rest are coloured blue. Then $G$ contains a blue clique $H$ of size $\lfloor k/3 \rfloor$ that can be found in polynomial time.*

*Proof.* We prove our claim by induction on $k$. Let $k \geq 3$ and consider the following cases:

(i) Assume the red degree of every vertex is at most 1; then we can obviously take half the vertices into $H$.

(ii) Assume the red degree of every vertex is exactly 2. Then the red subgraph consists of a number of cycles. From each cycle of size $t$, we can include every second vertex in $H$, hence obtaining at least $\lfloor t/2 \rfloor \geq t/3$ vertices ($t \geq 3$).

(iii) Otherwise $G$ has a vertex $u$ of red degree at most 1 and a vertex $v$ of red degree at least 2. If $u$ has red degree 1, let $w$ be its red neighbour, otherwise let $w$ be an arbitrary vertex (note that it might be $v = w$). Define $D := \{u, v, w\}$, $d := |D|$, and $G' := G - D$. Note that $d$ is 2 or 3 and $G'$ is a clique on $k - d$ vertices having at most $k - d$ red edges. By the induction hypothesis, we can find a blue clique $H'$ in $G'$ of size at least $\lfloor (k-d)/3 \rfloor$. But since $u$ has only blue edges to $G'$, we can add $u$ to $H'$ to obtain $H$ of size at least $\lfloor (k-d)/3 \rfloor + 1 \geq \lfloor k/3 \rfloor$. □

**Lemma 4.6.** *Given a (full) tree-web $\mathcal{W}$ of order $\ell$, one can construct a nice (full) tree-web $\mathcal{W}'$ of order at least $\lfloor \ell/2 \rfloor$ with $\mathcal{W}' \subseteq \mathcal{W}$ in polynomial time.*

*Proof.* Vertices of degree 1 are irrelevant to a tree-web and can be always removed; hence, we obtain and can always maintain property (1) of a nice tree-web. If a full subtree contains only one good vertex, we can extend the good path starting at that vertex to the root of the subtree and remove the rest of the subtree. Thus, we can always guarantee property (3), which implies properties (4) and (5). None of these operations changes the order of the tree-web.

It remains to show property (2). If a good path $P = v_0, \ldots, v_k$ intersects another path, i.e a path $Q \in \mathcal{Q}$ at vertex $v_1$, we colour $Q$ red. Since the number of good paths is at most $\ell^2$, we obtain at most $\ell^2$ red paths in $\mathcal{Q}$. Now, consider the subdivision $H$ of $\mathbf{K}_{\ell^2}$ that is contained in $\mathcal{I}(\mathcal{P}, \mathcal{Q})$; at most $\ell^2$ of the paths in $H$ that correspond to a subdivided edge of $\mathbf{K}_{\ell^2}$ contain a red vertex $Q \in \mathcal{Q}$; by Lemma 4.5, we can find a subdivision $H'$ of $\mathbf{K}_{\lfloor \ell^2/3 \rfloor}$ in $H$ that contains no red vertices and whose nails are a subset of the nails of $H$. Hence, by considering only the paths $\mathcal{P}' := \mathcal{P} \cap V(H')$ and $\mathcal{Q}' = \mathcal{Q} \cap V(H')$, we still have a subdivision of $\mathbf{K}_{\lfloor \ell^2/3 \rfloor}$ in $\mathcal{I}(\mathcal{P}', \mathcal{Q}')$. □

Using Theorem 4.2 (i) and Lemma 4.6, we can easily prove:

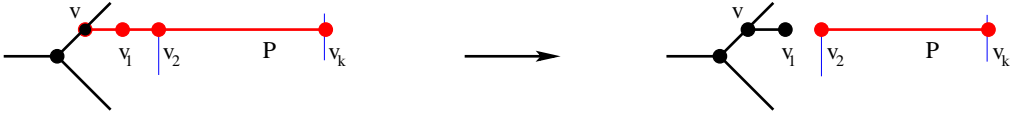

**Figure 3. Illustration of the operation** $\text{CUT}(v)$.

**Lemma 4.7.** *There is a constant $c$ and a polynomial-time algorithm that given a graph $G$ of treewidth at least $c\ell^{14}$ finds an $\ell \times \ell$-wall or a full nice tree-web of order $\ell$ in $G$.*

*Proof.* By choosing the constant $c$ appropriately, theorem 4.2 (i) implies that $G$ contains either $\mathbf{K}_{8\ell^2}$ as a minor or a topological grid-like minor of order $8\ell^2$. In the former case we are done, since an elementary $\ell \times \ell$-wall has maximum degree 3, is contained in $\mathbf{K}_{8\ell^2}$, and is thus a topological minor of $G$ implying that $G$ contains an $\ell \times \ell$-wall as a subgraph.

In the latter case, let $(T, (T_i)_{1 \leq i \leq t}, (A_i)_{1 \leq i \leq t}, B)$ be the $k$-web of order $t$ ($t = 3$ or $4$) and $(\mathcal{P}, \mathcal{Q})$ the topological grid-like minor returned by the algorithm. The theorem states that $\mathcal{P}$ is the set of paths connecting $T_1$ and $T_2$. Let $N$ be the set of nails in the model of $\mathbf{K}_{8\ell^2}$ in $\mathcal{I}(\mathcal{P}, \mathcal{Q})$. W.l.o.g. we may assume that $\mathcal{P}$ contains at least half of $N$; hence, by just considering the nails in $\mathcal{P}$, we still have a subdivision of $\mathbf{K}_{4\ell^2}$. By deleting $T_2, T_3$, and $T_4$ (if existent), we almost obtain a full tree-web of order $2\ell$, except that the paths in $\mathcal{P}$ that are not nails also intersect with $T_1$; but we can delete the first edge of these paths and obtain the desired full tree-web. The root of the tree can be chosen arbitrarily. Finally, we obtain our claim by appealing to Lemma 4.6. □

The following operation is essential for the proofs that follow.

**Definition 4.8.** Given a tree-web $\mathcal{W} = (G, T, r, A, \mathcal{P}, \mathcal{Q})$ and a good vertex $v \in A$, starting a path $P = vv_1 \ldots v_k$ of $\mathcal{P}$, the operation $\text{CUT}(v)$ is defined as removing the edge $v_1v_2$ from $G$, adding the edge $vv_1$ to $T$, removing the vertex $v$ from $A$, and iteratively removing vertices of degree 1 from $P$.

See Figure 3 for an illustration. Note that by starting with a nice tree-web, this operation does not affect the order of the tree-web. Next, we would like to identify a unique root for a tree-web:

**Definition 4.9.** A tree-web $\mathcal{W} = (G, T, r, A, \mathcal{P}, \mathcal{Q})$ *admits a definable root* if it contains exactly one vertex $r \in V(T)$ with $\deg_T(r) = 1$ and $\deg_G(r) = 3$ such that two components of $G - r$ are single vertices $s_1, s_2$ and the third contains at least one edge.

**Lemma 4.10.** *Given a full nice tree-web $\mathcal{W} = (G, T, r, A, \mathcal{P}, \mathcal{Q})$ with at least 3 good vertices, one can construct a subtree-web $\mathcal{W}' = (G', T', r', A', \mathcal{P}', \mathcal{Q}')$ of the same order in polynomial time such that $\mathcal{W}'$ is nice, admits a definable root, and $|A'| \geq |A|/3$.*

*Proof.* If $T$ has a vertex $v$ of degree 3, one of the components of $T - v$ contains at least $1/3$ of the good vertices; we prune the other two to become a single vertex each to obtain $\mathcal{W}'$ with root $v$. If, on the other hand, $T$ is a path, one of its endpoint $v_1$ is a good vertex connecting a path $P_1$ and its neighbour $v_2$ is a good vertex connecting a path $P_2$. By deleting the first edge of $P_1$ and applying the operation $\text{CUT}(v_2)$, $v_2$ becomes a definable root while losing only 2 good vertices. Since $\mathcal{W}$ is nice, these operations do not change the order of the tree-web; and by deleting redundant vertices of degree 1, we can make sure that $\mathcal{W}'$ is nice, too. □

### 4.3 Trees admitting a definable ordering

In this section we show how to prune a given rooted tree $T$ with maximum degree 3, so that there is an $\text{MSO}_2$-formula (not depending on $T$) which at each branching node of the tree distinguishes between the left and the right subtree. Assume we are given a subcubic tree $T$ with a root $r$ and a set $X$ of vertices of the tree marked as *good* and we would like to retain as many good vertices as possible. Throughout this section, $X$ will always denote the set of good vertices; and we assume $\deg_T(r) = 1$. We use the following notation (see Figure 4 for illustrations):

- If $v \in V(T)$ then the children of $v$ are all neighbours of $v$ not on the unique path from $v$ to $r$.
- A *leaf* of $T$ is a node of degree 1 in $T$, except $r$. A *good leaf* is a leaf that is good.
- A vertex is called *leafy* if it has degree 3 and is adjacent to a leaf.
- A *branching vertex* of $T$ is a vertex of degree 3 in $T$. A *proper branching vertex* is a branching vertex that is not leafy.
- An *artificial leaf* is a leaf that is not good and is adjacent to a branching vertex.
- Let $v \in V(T)$ be a vertex with child $u \in V(T)$ and $e = \{v, u\}$. The subtree $T_u$ of $T$ *rooted at $u$* is the component of $T - e$ containing $u$. The *extended subtree* of $u$ is defined as $T_u \cup e$.
- SUBTREE$_i(v)$ denotes the extended subtree of the $i^{\text{th}}$ child of $v$, where we number the children arbitrarily.
- CBV$(T, v)$ : closest branching vertex to $v$ in $T_v$; or the leaf of $T_v$ if $T_v$ is a path; is defined only if $v$ has degree 1 in $T_v$.
- $g_X(T) := |X \cap V(T)|$ is the number of good vertices in $T$. We omit the index $\cdot_X$ if it is clear from context.

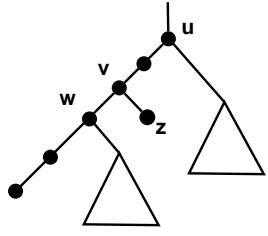

**Figure 4. The vertices $u$, $v$, and $w$ are properly marked; $v$ is leafy and $z$ an artificial leaf; $u$ and $w$ are proper branching vertices.**

**Definition 4.11.** Let $(T, r)$ be a rooted sub cubic tree;

- two vertices $u, v \in V(T)$ are *topological neighbours* if they are linked by a path whose inner vertices all have degree 2 in $T$;
- a branching vertex $v$ is called *properly marked* if it has a leaf or a leafy vertex as a topological neighbour in the subtree rooted at $v$; and
- $T$ is called *properly marked* if every branching vertex of $T$ is properly marked.

We now define a pruning algorithm PRUNE$(T, r)$ which, given a rooted subcubic tree $(T, r)$ outputs a tree $(T', r)$ that is properly marked (see Figure 5).

**Algorithm** PRUNE$(T, r)$.
*Input.* subcubic rooted tree $(T, r)$ with $\deg_T(r) = 1$.
*Output.* a properly marked subcubic rooted tree $(T', r)$ with $T' \subseteq T$.

If $T$ is a simple path than return $T$. Otherwise, let $v := \text{CBV}(T, r)$, $R$ be the path from $r$ to $v$, $T_1 := \text{SUBTREE}_1(v)$, and $T_2 := \text{SUBTREE}_2(v)$ with $g(T_1) \leq g(T_2)$.

1. If one of $T_1, T_2$ is a path, say $T_i$, return the tree obtained from $T$ by replacing $T_{3-i}$ by PRUNE$(T_{3-i})$.
2. Otherwise, let $u_1 := \text{CBV}(T_1, v)$. Let $T_{11} := \text{SUBTREE}_1(T_1, u_1)$ and $T_{12} := \text{SUBTREE}_2(T_1, u_1)$ with $g(T_{11}) \leq g(T_{12})$. Let $T_1'$ be the tree obtained from $T_1$ by cutting $T_{11}$ down to a single edge and replacing $T_{12}$ by $T_{12}' := \text{PRUNE}(T_{12}, u_1)$. Finally, return $T'$ as the union of $R$, $T_1'$, and $T_2' := \text{PRUNE}(T_2)$.

**Lemma 4.12.** *Let $(T, r)$ be a rooted subcubic tree and $X \subseteq V(T)$. $T$ contains a properly marked subtree $T'$ such that $g_X(T') \geq g_X(T)^{\frac{2}{3}}$. Furthermore, $T'$ can be computed in polynomial time on input $(T, r)$.*

*Proof.* Let $T' := \text{PRUNE}(T, r)$. We claim that $(T', r)$ fulfils the requirements of the lemma. We prove the claim by induction on the order $n := |T|$ of $T$. If $T$ is a path there is nothing to show. Otherwise, the fact that $T'$ is properly marked is immediate from our recursive construction by induction. It remains to bound the number of good vertices that remain after the pruning. We first observe that for all $\frac{1}{2} \leq \beta \leq 1$

$$\left(\frac{1-\beta}{2}\right)^{\frac{2}{3}} + \beta^{\frac{2}{3}} \geq 1. \tag{1}$$

If $q, q_1, q_2$ are non-negative integers with $q = q_1 + q_2$ and $q_1 \leq q_2$, we have $q_2 = \beta q$ and $q_1 = (1-\beta)q$, for some $\beta \geq \frac{1}{2}$. Hence, we obtain with Inequality (1)

$$\begin{aligned} q_1^{\frac{2}{3}} + q_2 \geq q_1 + q_2^{\frac{2}{3}} \geq q_1^{\frac{2}{3}} + q_2^{\frac{2}{3}} &\geq \left(\frac{q_1}{2}\right)^{\frac{2}{3}} + q_2^{\frac{2}{3}} \\ &= q^{\frac{2}{3}} \cdot \left(\left(\frac{1-\beta}{2}\right)^{\frac{2}{3}} + \beta^{\frac{2}{3}}\right) \\ &\geq q^{\frac{2}{3}} = (q_1 + q_2)^{\frac{2}{3}}. \end{aligned} \tag{2}$$

Let $v$, $R$, $T_1$, and $T_2$ be defined as in the algorithm. Define $q_0 := g(R - v)$, $q_1 := g(T_1 - v)$, $q_2 := g(T_2)$, $q := q_1 + q_2$, and $q' := g(T_v')$. First, note that it suffices to show $q' \geq q^{\frac{2}{3}}$ since this implies

$$g(T') = q_0 + q' \geq q_0 + q^{\frac{2}{3}} \geq (q_0 + q)^{\frac{2}{3}} = g(T)^{\frac{2}{3}}$$

by Inequality (2). Consider the following cases:

(i) If $T_1$ is a path, then $q' \geq q_1 + q_2^{\frac{2}{3}} \geq q^{\frac{2}{3}}$ by Inequality (2). Similarly, if $T_2$ is a path, then $q' \geq q_1^{\frac{2}{3}} + q_2 \geq q^{\frac{2}{3}}$.

(ii) Otherwise, let $T_{12}'$ and $T_2'$ be defined as in Step 2 of the algorithm and let $q_2' := g(T_2')$ and $q_{12}' := g(T_{12}')$. Furthermore, let $P$ be the path from $u$ to $v$ excluding $u$ and $v$ and let $q_P := g(P)$. Using Inequality (2) twice more, we obtain

$$\begin{aligned} q' = q_P + q_{12}' + q_2' &\geq q_P + \left(\frac{q_1 - q_P}{2}\right)^{\frac{2}{3}} + q_2^{\frac{2}{3}} \\ &\geq \left(\frac{q_1}{2}\right)^{\frac{2}{3}} + q_2^{\frac{2}{3}} \geq q^{\frac{2}{3}}. \end{aligned}$$

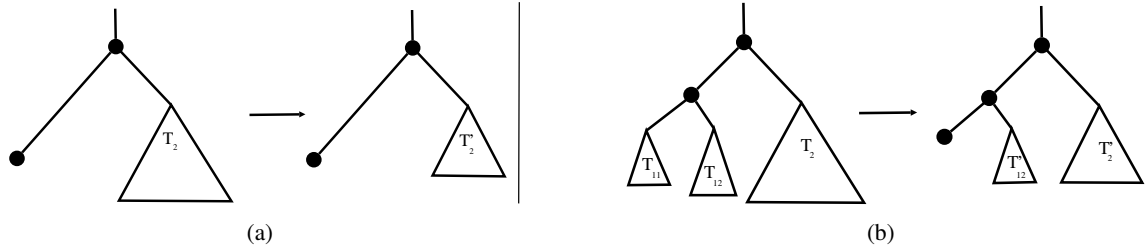

**Figure 5. (a) Case (1) of algorithm** PRUNE**; (b) case (2) of algorithm** PRUNE**.**

□

Once we have a properly marked tree, it is possible to identify left and right subtrees in a proper way using parity considerations, as follows.

**Definition 4.13.** Let $(T, r)$ be a subcubic tree rooted at vertex $r$ of degree 1 and $X$ a set of good vertices that lies flat in $T$. We say the tuple $(T, r, X)$ *admits a definable order* if all leaves are artificial or good and for all branching vertices $v$ with extended subtrees $T_i := \text{SUBTREE}_i(v)$, for $i = 1, 2$, exactly one of the following is true. Along with the following conditions we will label some subtrees as *left* and others as *right*.

1. At least one of $T_1, T_2$ is a single edge, say $T_1$. If $T_2$ is also a single edge, then exactly one of $T_1$ and $T_2$ contains a good leaf, say $T_1$; in either case, $T_1$ is left and $T_2$ is right.
2. Exactly one of $T_1, T_2$ is a simple path, say $T_1$. Then $T_1$ is left and $T_2$ is right.
3. Let $u_i$ be the closest proper branching vertex to $v$ in $T_i$ if one exists; otherwise let $u_i$ be the good leaf of $T_i$ farthest away from $v$. Let $P_i$ the path connecting $v$ and $u_i$ in $T_i$. We define $g_i$ to be the number of vertices on $P_i - v$ that are *good or leafy*. We require that exactly one of $g_1, g_2$ is odd, say $g_1$; then $T_1$ is left and $T_2$ is right.

The *canonical order* $\leq_T$ of $(T, r)$ is defined as follows. Let $x \neq y \in V(T)$ and let $v$ be the closest common ancestor of $x, y$. Then $x \leq_T y$ if and only if $v = x$ or $x$ is in the left subtree of $v$ and $y$ in the right.

**Lemma 4.14.** *Let $(T, r)$ be a subcubic tree rooted at vertex $r$ of degree 1 and $X \subseteq V(T)$ a given set of good vertices that lies flat in $T$. $T$ contains a subtree $T'$ and a set $X' \subseteq X \cap T'$ with $|X'| \geq |X|^{\frac{2}{3}}/2$ such that $(T', r, X')$ admits a definable order and $X'$ is* totally *ordered by the canonical order $\leq_{T'}$. Furthermore, $T'$ can be computed in polynomial time.*

*Proof.* W.l.o.g. we assume all the leaves of $T$ are good; otherwise we go from $T$ to the smallest subtree of $T$ containing the root and all good vertices; hence, all the leaves of $T$ are good leaves. Then we apply Lemma 4.12 to obtain a properly marked subtree $T''$ of $T$ and a set $X'' := X \cap T''$ with $|X''| \geq |X|^{\frac{2}{3}}$. Note that when counting the number of good vertices of $T''$ in Lemma 4.12, we do not consider the leaf that replaces a subtree in step (2) of algorithm PRUNE a good vertex, even though it might happen to be one; hence, we consider all these leaves artificial leaves and are free to remove them without losing good vertices. All other leaves are still good vertices. Now we consider each branching vertex $v$ of $T''$ in a bottom-up fashion, i.e. in a post-order traversal of the tree, and consider the following cases; let $T_i, u_i$, and $g_i$ be defined as in Definition 4.13:

(i) If both of $T_1, T_2$ are single edges, ignore $v$.
(ii) Suppose $T_1$ and $T_2$ are both simple paths of length at least 2 and $g_1$ and $g_2$ have the same parity. Then $u_1$ and $u_2$ are both good leaves. If there is no other good vertex in $T_1$, we cut $T_1$ down to a single edge, i.e. make it an artificial leaf. Otherwise, let $w$ be the good vertex closest to $u_1$ and replace $T_1$ by the path from $v$ to $w$. In either case, we lose exactly one good vertex in the subtree rooted at $v$; and our construction ensures that case (ii) does not occur for any ancestor of $v$ in the tree.

(iii) Otherwise, if at least one of $T_1, T_2$ is a simple path, ignore $v$.

(iv) Otherwise $v$ is a proper branching vertex and Lemma 4.12 guarantees that $v$ has a leafy vertex, adjacent to an artificial leaf $w$, as a topological neighbour in one of its subtrees. If $g_1$ and $g_2$ are of the same parity, we simply remove $w$ and obtain our desired property without losing any good vertex.

We let $T'$ be the tree obtained after the traversal above is finished and $X' := T' \cap X - D$, where $D$ contains one of every two good leaves that are siblings as in case (i). Then, it is evident by our construction that $(T', r, X')$ admits a definable order and all leaves are either good or artificial. Furthermore, $X'$ contains at least half the vertices of $X''$, since the subtrees on which case (i) or (ii) apply are all disjoint and at most half of their good vertices are not included in $X'$. □

### 4.4 Tree-Ordered Webs

We show how to prune the tree $T$ of a given nice full tree-web $(G, T, r, A, \mathcal{P}, \mathcal{Q})$, so that there is an MSO$_2$-formula which can detect the nodes of $T$ in $G$ and at each branching node of the tree distinguishes between the left and right subtree.

**Definition 4.15.** Let $\mathcal{W} = (G, T, r, A, \mathcal{P}, \mathcal{Q})$ be a tree-web:

- a *leaf-mark* of $\mathcal{W}$ is a path $v_1 v_2 v_3$ in $G$ such that $v_1$ is a good leaf of $T$, $v_2$ is of degree $2$ in $G$ and $v_3$ is of degree $1$ in $G$ (see Figure 6);
- we use the notions *topological neighbour* and *properly marked* with respect to *degrees in $G$* (as opposed to degrees in $T$ in the previous subsection; this does make an important difference, as good vertices have degree $2$ in $T$ but degree $3$ in $G$);
- a vertex $v \in V(G)$ is *special in $G$* if it has degree $3$ or $4$ and is *not* properly marked, i.e. does not have a leaf or a leafy vertex as a topological neighbour;
- we let $spec(G)$ denote the set of special vertices in $G$.

**Definition 4.16.** A *tree-ordered web* of order $\ell$ is a tuple $(G, T, r, A, \mathcal{P}, \mathcal{Q})$ such that

1. $(G, T, r, A, \mathcal{P}, \mathcal{Q})$ is a tree-web of order $\ell$ admitting the definable root $r$;
2. $A$ is the set of vertices of degree $3$ in $G$ not in $spec(G)$ but having a topological neighbour in $spec(G)$;
3. $T$ is contained in the component of $G - spec(G)$ containing $r$;
4. every leaf of $T$ is either artificial or incident to a leaf-mark;
5. $(T, r, A)$ admits a definable order;
6. $G$ consists only of $T \cup \bigcup \mathcal{P} \cup \bigcup \mathcal{Q}$, the marking of $r$, and the leaf-marks; and
7. no vertex of $V(\bigcup \mathcal{P}) \cup V(\bigcup \mathcal{Q})$ has degree $1$ in $G$.

A main ingredient of the proof of Lemma 4.17 below is the observation that Definition 4.13 allows us to cut a good path and make it an artificial leaf without destroying the definable order; this is because in the third case of Definition 4.13, we consider vertices that are *good or leafy* and the operation CUT only turns a good vertex into a leafy vertex. Hence, we can cut away about every second good path to ensure that vertices of the tree do not land in $spec(G)$. We also observe that if the number of the leaves of the tree is large enough, we can just keep the good paths starting at the leaves; and otherwise, the number of proper branching vertices is small and we do not need to cut away too many good paths.

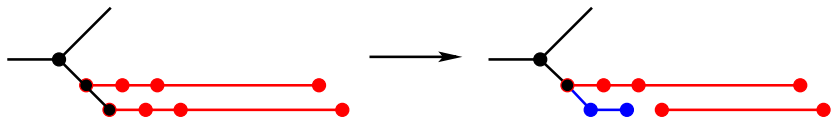

**Figure 6.** Turning a good leaf with a good neighbour into a leaf with a leaf-mark.

**Lemma 4.17.** *There exists a constant $c$ such that if $\mathcal{W}_0 = (G_0, T_0, r_0, A_0, \mathcal{P}_0, \mathcal{Q}_0)$ is a given nice full tree-web of order $c\ell$, then there exists a tree-web $\mathcal{W} = (G, T, r, A, \mathcal{P}, \mathcal{Q})$ with $\mathcal{W} \subseteq \mathcal{W}_0$ and a vertex $r \in V(G)$ such that $(G, T, r, A, \mathcal{P}, \mathcal{Q})$ is a tree-ordered web of order $\ell$ with $|A| \geq 15\ell$; furthermore, $\mathcal{W}$ can be computed in polynomial time.*

*Proof.* First, we apply Lemma 4.10 to obtain a nice tree-web $\mathcal{W}_1 = (G_1, T_1, r, A_1, \mathcal{P}_1, \mathcal{Q}_1)$ with definable root $r$. The lemma guarantees that $|A_1| \geq |A_0|/3$. Recall that this way, $r$ is of degree 1 in $T$ but is never considered a leaf.

Since $\mathcal{W}_1$ is nice, every leaf of $T_1$ is good and is adjacent to another good vertex of $T_1$. Let $T_1' \subseteq T_1$ be the subtree of $T_1$ obtained by removing all the leaves of $T_1$. Note that all the leaves of $T_1'$ are still good. Let $A_1' = A_1 \cap V(T_1')$ and observe that $|A_1'| \geq |A_1|/2$. Next, we invoke Lemma 4.14 on $(T_1', r)$ and $A_1'$ to obtain a tuple $(T_2, r, A_2)$ admitting a definable order with $|A_2| \geq |A_1'|^{\frac{2}{3}}/2$. The lemma guarantees that every leaf of $T_2$ is either artificial or good.

Let $v$ be a good leaf of $T_2$. By our construction above, there must exist an edge $vu \in E(T_1) - E(T_2)$. Now (i) if $u$ is not a good vertex of $T_1$, then $u$ cannot be a leaf of $T_1$, and hence there must exist another edge $uw \in E(T_1) - E(T_2)$ with $w \neq v$; in this case, $vuw$ is a leaf-mark for $v$; (ii) otherwise, $u$ is a good vertex of $T_1$ starting a path $P = uw_1w_2\ldots$; we delete the edge $w_1w_2$ as in the CUT operation to obtain the leaf-mark $vuw_1$ for $v$ (see Figure 6); since we started with a nice tree-web, this operation does not change the order of the tree-web.

We apply the procedure above to every good leaf of $T_2$, remove all edges of $T_1$ from $G$ that do not appear in $T_2$ or in leaf-marks, and iteratively remove redundant vertices of degree 1 appearing in $(\mathcal{P}, \mathcal{Q})$. Let $\mathcal{W}_2 = (G_2, T_2, r, A_2, \mathcal{P}_2, \mathcal{Q}_2)$ be the resulting tree-web.

Let $b$ be the number of proper branching vertices of $T_2$ and $t$ the number of good leaves. Since $T_2$ is subcubic, we have $b \leq t$. We obtain the tree-web $\mathcal{W}_3 = (G_3, T_3, r, A_3, \mathcal{P}_3, \mathcal{Q}_3)$ as follows:

(i) If $t \geq |A_2|/4$, we let $A_3$ be the set of good leaves of $T_2$ and obtain $\mathcal{W}_3$ by performing the operation CUT on every good vertex not in $A_3$.
(ii) Otherwise, consider each proper branching vertex $v$ of $T_2$ and let $P_1, P_2$ be the paths from $v$ to the closest proper branching vertex or good leaf of each of the two extended subtrees of $v$. Let $u_1, \ldots, u_p$ be the good vertices on $P_1$, in this order, and similarly, $w_1, \ldots, w_q$ the good vertices on $P_2$. If one of $p, q$ is 1, we assume w.l.o.g. that $p = 1$; if both are 1, then we let $w_1, \ldots, w_q$ belong to the subtree that contains an artificial leaf (note that there is such a subtree, since the parities of the total number of good or leafy vertices must be different). Apply the operation $\text{CUT}(u_i)$ for every odd $1 \leq i \leq p$ and the operation $\text{CUT}(w_i)$ for every even $1 \leq j \leq q$. This way, it is guaranteed that $v$ retains a leafy vertex as a topological neighbour and still, at least $\lfloor (p+q)/2 \rfloor$ good vertices on $P_1 \cup P_2$ are left. If $p + q$ is odd, we charge one unit of penalty to $v$. Let $T_3$ and $A_3$ be the tree and good vertices after this operation is performed on every proper branching vertex. Since every proper branching vertex is charged to at most once, the number of good vertices that remain is at least $|A_2|/2 - b$; but we have $b \leq t \leq |A_2|/4$, and hence we obtain $|A_3| \geq |A_2|/4$.

We claim that $\mathcal{W} := \mathcal{W}_3$ is the desired tree-ordered web specified in the lemma. Indeed, note that since we started with a nice tree-web, none of the operations above changed the order of the tree-webs we worked with; also every branching vertex and every good vertex in $T$ is properly marked by a leafy topological neighbour while vertices of $\mathcal{P} \cup \mathcal{Q}$ do not have this property and thus belong to $spec(G)$; furthermore, $(T_3, r, A_3)$ admits the same definable order as $(T_2, r, A_2)$ because the CUT operation only changes good vertices into leafy vertices, which does not make a difference in the definable order. Hence, all the properties of Definition 4.16 are fulfilled. Finally, recall that $|A_0| = c^2 \ell^2$; we have $|A_3| \geq \frac{|A_0|^{\frac{2}{3}}}{27} = \frac{c^{\frac{4}{3}}}{27} \cdot \ell^{\frac{4}{3}}$, and so $|A_3| \geq 15\ell$ is also fulfilled if the constant $c$ is large enough (if $\ell$ is larger than a constant, then $c = 1$; otherwise $c \leq 91$ suffices). □

## 4.5 Labelling Tree-Ordered Webs

We will show next how to encode a word $w := w_1 \ldots w_t \in \{0,1\}^\star$ in a tree-ordered web of order $2t$. We first need the following simple combinatorial lemma.

**Lemma 4.18.** *Let $G$ be a directed graph on $k$ vertices with maximum outdegree $d$. Then $G$ contains an independent set of size $\left\lceil \frac{k}{2d+1} \right\rceil$ which can be computed in polynomial time.*

*Proof.* As the maximal outdegree of each vertex is at most $d$, the graph contains at most $kd$ edges, i.e. in the underlying undirected graph, the sum of the vertex degrees is at most $2kd$. Hence, there is a vertex of total degree at most $2d$. We can add it to the independent set and remove all its in- and out-neighbours. Proceeding in this way we find an independent set of size $\left\lceil \frac{k}{2d+1} \right\rceil$. □

A *single cross* is a subcubic tree with four leaves having the shape depicted in Figure 7 (a); a *double cross* is a subcubic tree with five leaves having the shape depicted in Figure 7 (b) (where the dashed lines indicate paths). The right-most vertex of each cross, as drawn in Figure 7, is called the *base* of the cross.

**Definition 4.19.** A *labelled tree-ordered web of order $\ell$ and length $k$* is a tuple $\mathcal{W} := (G, T, r, A, \mathcal{P}, \mathcal{Q}, X, C)$ where

1. $(G[V(G - C) \cup X], T, r, A, \mathcal{P}, \mathcal{Q})$ is a tree-ordered web of order $\ell$ except that we require $(T, r, A \cup X)$ to admit a definable order instead of $(T, r, A)$,
2. the root $r$ does not have a leafy vertex as a topological neighbour,
3. $C$ is a set of disjoint single and double crosses,
4. $X = V(T) \cap V(C)$ is the set of bases of the crosses in $C$ and lies flat in $T$,
5. $|X| = |A| = k$,
6. if $X = \{x_1, \ldots, x_k\}$ and $A = \{v_1, \ldots, v_k\}$ then $x_1 \leq_T v_1 \leq_T x_2 \cdots \leq_T x_k \leq_T v_k$.

The word *encoded by* $\mathcal{W}$ is $w := w_1 \ldots w_k \in \{0,1\}^k$ with $w_i := 0$ if $x_i$ is the base of a single cross in $C$ and $w_i := 1$ if $x_i$ is the base of a double cross of $C$. $\mathcal{W}$ is called *configurable* if $C$ consists only of double crosses.

A labelled tree-ordered web encoding the word 010 is indicated in Figure 2.

**Lemma 4.20.** *For $\ell \geq 3$, let $\mathcal{W} = (G, T, r, A, \mathcal{P}, \mathcal{Q})$ be a given tree-ordered web of order $2\ell$ with $|A| \geq 30\ell$. There exists a configurable labelled tree-ordered web $\mathcal{W}' = (G', T', r', A', \mathcal{P}', \mathcal{Q}', X', C')$ of order $\ell$ and length $\ell$ with $G' \subseteq G$ that can be computed in polynomial time.*

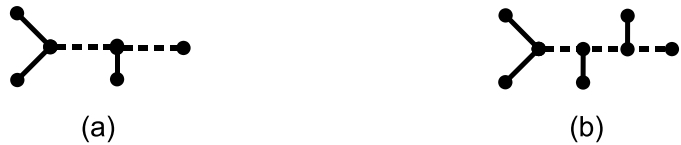

**Figure 7.** A (a) single and a (b) double cross.

*Proof.* First, note that any good path $P \in \mathcal{P}$ can be easily transformed to a double cross: since $P$ is a nail of the grid-like minor $(\mathcal{P}, \mathcal{Q})$ of order $4\ell^2$, $P$ intersects with at least 3 paths of $\mathcal{Q}$ if $\ell \geq 1$. The first 3 of these paths can be cut in a way to create one of the double crosses depicted in Fig. 7 (b). By doing so, we could destroy at most 7 other paths: 3 paths of $\mathcal{Q}$ and 4 paths of $\mathcal{P}$ that might have intersected the 4 leaves of the double cross (the base is part of the tree $T$). Each path $R \in \mathcal{P} \cup \mathcal{Q}$ that is not a nail might be used on at most one subdivided edge connecting the nails $P_1, P_2 \in \mathcal{P}$ in the image of $\mathbf{K}_{4\ell^2}$ in $\mathcal{I}(\mathcal{P}, \mathcal{Q})$. Assign $R$ arbitrarily, say, to $P_1$. If $R$ is destroyed by building a cross, we consider $P_1$ being destroyed, instead.

Consider a digraph $\mathcal{D}$ having a vertex $u_P$ for each good path in $\mathcal{P}$ and a directed edge from $u_P$ to $u_{P'}$ if turning $P$ into a double cross destroys $P'$. The maximum outdegree of this digraph is 7 and hence, by Lemma 4.18, there exists a set $Y_0 \subseteq A$ of size at least $\frac{|A|}{15} \geq 2\ell$ of good vertices such that the vertices in $\mathcal{D}$ that correspond to the good paths starting at $Y_0$ form an independent set in $\mathcal{D}$.

Let $Y := \{y_1, \ldots, y_{2\ell}\}$ be a subset of exactly $2\ell$ vertices of $Y_0$ such that $y_1 \leq_T y_2 \leq_T \cdots \leq_T y_{2\ell}$. We define $X' := \{y_i \mid i \text{ is odd}\}$ and $A' := \{y_i \mid i \text{ is even}\}$. We transform every good path starting at a vertex in $X'$ into a double cross and remove all the paths that get destroyed. Let $C'$ be the set of double crosses we obtain this way. We also perform the operation CUT$(v)$ on every good vertex that is not in $Y$. If the root $r$ has a leafy vertex with leaf $w$ as a topological neighbour, we remove $w$ and repeat this process, if necessary; note that such leaves are irrelevant to the tree-order and can be safely removed. Finally, we repeatedly remove all redundant vertices of degree 1 of $(\mathcal{P}, \mathcal{Q})$.

Let $\mathcal{P}'$ and $\mathcal{Q}'$ be the (parts of) the paths that remain, $T'$ the tree obtained from $T$ after the cut operations, $r' := r$, and $G'$ be the union of $T' \cup C' \cup \mathcal{P}' \cup \mathcal{Q}'$ with the marking of the root and the leaf-marks. We claim $\mathcal{W}' = (G', T', r', A', \mathcal{P}', \mathcal{Q}', X', C')$ is the desired configurable labelled tree-ordered web of order $\ell$ and length $\ell$.

The fact that $(T', r', A' \cup X')$ admits a definable order follows on one hand, from the observation that Definition 4.13 allows turning a good vertex into a leafy vertex by the CUT operation without changing the canonical order of the tree; and on the other hand, from the fact that the good vertices that started good paths that are now turned into crosses are now in $X$ and thus still count as good vertices. Hence, the canonical order of $(T', r', A' \cup X')$ is indeed the same as the canonical order of $(T, r, A)$. The number of destroyed paths is at most $8|X'| = 8\ell$, i.e. we lose at most $8\ell$ nails of the subdivision of $\mathbf{K}_{4\ell^2}$ in $\mathcal{I}(\mathcal{P}, \mathcal{Q})$; hence $\mathcal{I}(\mathcal{P}', \mathcal{Q}')$ still contains a subdivision of $\mathbf{K}_{\ell^2}$ if $\ell \geq 3$. All other requirements of Definition 4.19 are immediate from our construction. □

The definition below is needed in Section 6:

**Definition 4.21.** If $\mathcal{W}$ is a labelled tree-ordered web of order $\ell^d$ and length $\ell$ encoding a word $w = w_1 \ldots w_\ell$, we say that $\mathcal{W}$ *encodes $w$ with power $d$*.

Theorem 4.22 sums up the main algorithmic results of this work: :

**Theorem 4.22.** *Let a word $w = w_1 \ldots w_\ell \in \{0, 1\}^\star$, a graph $G$, and an integer $d$ be given. There is a constant $c$ such that if the treewidth of $G$ is at least $c\ell^{14d}$ then $G$ contains either an $\ell^d \times \ell^d$-wall or a labelled tree-ordered web $\mathcal{W}$ that encodes $w$ with power $d$. Furthermore, either outcome can be computed in polynomial time.*

*Proof.* By applying Lemma 4.7 to $G$, we obtain either an $\ell^d \times \ell^d$-wall or a full nice tree-web of order $c'\ell^d$, for a suitable constant $c'$ if $c$ is chosen appropriately. In the former case, we are done and in the latter case, we invoke Lemma 4.17 and Lemma 4.20 in order and obtain a configurable labelled tree-ordered web $\mathcal{W}' = (G', T', r', A', \mathcal{P}', \mathcal{Q}', X', C')$ of order $\ell^d$ and length $\ell^d$. We apply the operation CUT$(v)$ to all but the first $\ell$ good paths in $A'$, remove all but the first $\ell$ double crosses in $C$, and cut some double crosses to single crosses according to $w$. The labelled tree-ordered web $\mathcal{W}$ that remains fulfils our requirements. Since all the Lemmas that we used require only polynomial time, the whole procedure does, too. □

# 5 Defining a Labelled Tree-Ordered Web in MSO$_2$

In this section, we aim at defining the various parts of a labelled tree-ordered web in MSO$_2$. We start by stating some auxiliary formulas that we need in our construction of the main formulas. We make use of the notation and basic formulas introduced in Section 3 using the incidence structure encoding of graphs.

- *adj*$(v, w, H) := v \neq w \wedge \exists e \in H (v \in e \wedge w \in e)$ says $v$ and $w$ are adjacent in $H$;
- *pathends*$(v, w, P) := path(P) \wedge deg^{=1}(v, P) \wedge deg^{=1}(w, P)$ says $P$ is a path with endpoints $v$ and $w$;
- *topneigh*$(v, w, P, H) := P \subseteq H \wedge pathends(v, w, P) \wedge \forall z \in V(P)(z \neq v \wedge z \neq w \rightarrow deg^{=2}(z, H))$ says $u$ and $w$ are topological neighbours connected by $P$ in $H$;
- *leaf*$(v, T) := deg^{=1}(v, T)$ says $v$ is a leaf of $T$;
- *leafy*$(v, T) := deg^{=3}(v, T) \wedge \exists w (adj(v, w, T) \wedge leaf(w, T))$ says $v$ is leafy in $T$;
- *pbranch*$(v, T) := deg^{=3}(v, T) \wedge \neg leafy(v, T)$ says $v$ is a proper branching vertex;

Henceforth, we assume $T$ is a tree and that a formula *root*$(r, T)$ is given.

- *ancstr*$(v, a, T) := \exists r \big(root(r, T) \wedge (r = a \vee \exists P \subseteq T (pathends(v, r, P) \wedge a \in V(P)))\big)$ says $a$ is an ancestor of $v$ in $T$;
- *cca*$(v, w, a, T) := ancstr(v, a, T) \wedge ancstr(w, a, T) \wedge \neg \exists a' \big(a \neq a' \wedge ancstr(v, a', T) \wedge ancstr(w, a', T) \wedge ancstr(a', a, T)\big)$ says $a$ is the closest common ancestor of $v$ and $w$ in $T$;
- *parent*$(v, p, T) := ancstr(v, p, T) \wedge adj(v, p, T)$ says $p$ is the parent of $v$ in $T$;
- *child*$(v, c, T) := adj(v, c, T) \wedge \neg parent(v, c, T)$ says $c$ is a child of $v$ in $T$;
- *subtree*$(v, H, T) := H \subseteq T \wedge v \in V(H) \wedge \forall p (parent(v, p, T) \rightarrow p \notin V(H) \wedge \forall e \in T(p \notin e \wedge e \cap V(H) \neq \varnothing \rightarrow e \in H)$ expresses for any vertex $v$ other than the root of $T$ that $H$ is the subtree of $T$ rooted at $v$.
- *extsubtree*$(v, H, T) := \exists c, e, H' \, child(v, c, T) \wedge subtree(c, H', T) \wedge c \in e \wedge v \in e \wedge H = H' \cup e$ says $H$ is the extended subtree of a child of $v$ in $T$;

**Lemma 5.1.** *There exists a uniform* MSO$_2$-*formula* $\varphi_\preceq(x, y, T)$ *which defines the canonical order* $\leq_T$ *on any rooted subcubic tree* $(T, r)$ *and set* $X \subseteq V(T)$ *in a graph* $G$, *assuming that* $(T, r, X)$ *admits a definable order and that* MSO$_2$-*formulas* $\varphi_R(v, T)$ *and* $\varphi_X(v)$ *defining the root of* $T$ *and the set* $X$, *respectively, are given.*

*Proof.* We have to define the conditions of Definition 4.13 in MSO$_2$. The first two conditions are easily captured by the following formula:

$$left_1(T_1, T_2) := \big(\exists^{=1} e(e \in T_1) \wedge \exists^{\geq 2} e(e \in T_2)\big) \vee \big(path(T_1) \wedge \neg path(T_2)\big) \vee$$
$$\big(\exists^{=1} e(e \in T_1 \wedge \exists v(v \in e \wedge \varphi_X(v))) \wedge \exists^{=1} e(e \in T_2)\big)$$

For two subgraphs $T_1$ and $T_2$, this formula says that if $T_1$ is a single edge but $T_2$ is not, or if $T_1$ is a simple path and $T_2$ is not, or $T_1$ and $T_2$ are both single edges and $T_1$ contains a good vertex, then $T_1$ is left of $T_2$. To capture the third condition, we need to compare the parity of good or leafy vertices on a path. So, let $gl(v) := \varphi_X(v) \vee leafy(v)$. First, we define an auxiliary formula expressing that a vertex $v$ has a topological neighbour $w$ such that none of the internal vertices of the subpath from $v$ to $w$ are good or leafy:

$$topneigh_{gl}(v, w, T) :=$$
$$\exists P \subseteq T \big(pathends(v, w, P) \wedge \neg \exists z \in V(P) \, (z \neq v \wedge z \neq w \wedge gl(z))\big)$$

Consider the following formula that guarantees that a path $P$ from $v$ to $w$ contains an odd number of good or leafy vertices, where we assume $v$ is neither good nor leafy:

$$\begin{aligned}
odd_{gl}(P, v, w) := {}& \exists C_1, C_2 \subseteq V(P) \\
& \bigl(C_1 \cap C_2 = \emptyset \wedge \forall x \in C_1 \cup C_2\, gl(x) \wedge \forall x \in V(P)(x \neq v \wedge gl(x) \to x \in C_1 \cup C_2) \wedge \\
& \forall x \in C_2 \exists y_1, y_2 \in C_1 \bigl(y_1 \neq y_2 \wedge topneigh_{gl}(x, y_1, P) \wedge topneigh_{gl}(x, y_2, P)\bigr) \wedge \\
& \forall x \in C_1 \exists y_1, y_2 \bigl((y_1 = v \vee y_1 \in C_2) \wedge (y_2 = w \vee y_2 \in C_2) \wedge y_1 \neq y_2 \wedge \\
& \qquad topneigh_{gl}(x, y_1, P) \wedge (topneigh_{gl}(x, y_2, P) \vee x = w)\bigr) \wedge \\
& v \notin C_1 \cup C_2 \wedge \exists x \in C_1\, topneigh_{gl}(v, x)\bigr)
\end{aligned}$$

The idea is to colour the good or leafy vertices on $P$ with two colours $C_1$ and $C_2$, such that a vertex of one colour has only vertices of the other colour as a direct topological neighbour. Now if we guarantee that the first and last vertex are coloured $C_1$, we have an odd number of good or leafy vertices on $P$. The next formula says that $w$ is the closest proper branching vertex to $v$ in a given subtree $H$ connected to $v$ by the path $P$:

$$\begin{aligned}
closest\text{-}pbv(v, w, P, H) := {}& pathends(v, w, P) \wedge pbranch(w, H) \wedge \\
& \neg \exists z \in V(P) \bigl(z \neq v \wedge z \neq w \wedge pbranch(z, H)\bigr)
\end{aligned}$$

Similarly, we can define the property that $w$ is the farthest good leaf from $v$ if no proper branching vertex occurs in $H$:

$$\begin{aligned}
farthest\text{-}leaf(v, w, P, H) := {}& pathends(v, w, P) \wedge leaf(w, H) \wedge \varphi_X(w) \wedge \\
& \forall z \in V(P) \bigl(z \neq v \wedge deg^{=3}(z, H) \to \exists y (y \neq w \wedge leaf(y, H) \wedge adj(z, y, H))\bigr)
\end{aligned}$$

Now if $T_1$ and $T_2$ are the extended subtrees of the children of a branching vertex $v$, we can determine if $T_1$ is left of $T_2$ according to Definition 4.13 as follows:

$$\begin{aligned}
left(v, T_1, T_2) := {}& left_1(T_1, T_2) \vee \bigl(\neg left_1(T_2, T_1) \wedge \\
& \exists w, P \subseteq T_1 \bigl((closest\text{-}pbv(v, w, P, T_1) \vee farthest\text{-}leaf(v, w, P, T_1)) \wedge odd_{gl}(P, v, w)\bigr)\bigr)
\end{aligned}$$

Finally, we can define the canonical order $\leq_T$:

$$\begin{aligned}
\varphi_\leq(x, y, T) := {}& \exists v \bigl(cca(x, y, v, T) \wedge \bigl(v = x \vee \bigl(v \neq y \wedge \\
& \exists T_1, T_2 \subseteq T\, (extsubtree(v, T_1, T) \wedge extsubtree(v, T_2, T) \wedge \\
& x \in V(T_1) \wedge y \in V(T_2) \wedge left(T_1, T_2))\bigr)\bigr)\bigr)
\end{aligned}$$

$\square$

**Lemma 5.2.** *Given a labelled tree-ordered web $\mathcal{W} = (G, T, r, A, \mathcal{P}, \mathcal{Q}, X, C)$, there exist $\mathrm{MSO}_2$-formulas $\varphi_T(H)$, $\varphi_R(v)$, $\varphi_A(v)$, $\varphi_X(v)$, $\varphi_{Cr_1}(H)$, $\varphi_{Cr_2}(H)$, $\varphi_{PQ}(H)$, and $\varphi_\preceq(x, y)$ defining the tree $T$, its root $r$, the set of good vertices $A$, the bases of the crosses $X$, the single and double crosses $C$, the edge set of the grid-like minor $\bigcup \mathcal{P} \cup \bigcup \mathcal{Q}$, and the canonical order $\leq_T$, respectively. These formulas are uniform and do not depend on $\mathcal{W}$ in any form.*

*Proof.* We construct the required formulas gradually, occasionally using auxiliary formulas; for formulas that we defined previously as $\varphi(\cdot, H)$, we sometimes write simply $\varphi(\cdot)$ for $\varphi(\cdot, E)$:

- *rootish*$(v) := deg^{=3}(v) \wedge \exists x, y\, (leaf(x) \wedge leaf(y) \wedge adj(v, x) \wedge adj(v, y))$ says that $v$ is of degree 3 and has two neighbours of degree 1;

- $\varphi_R(v) := \textit{rootish}(v) \wedge \exists w, P\, \big(\textit{topneigh}(v, w, P) \wedge \textit{pbranch}(w)\big)$ uniquely defines the root of the tree $T$;

- $\textit{leafytopneigh}(v, w, P) := \textit{topneigh}(v, w, P) \wedge \textit{leafy}(w)$ says that $w$ is leafy topological neighbour of $v$ connected by path $P$;

- $\textit{cross}_1(x, y, b, P_1, P_2) := \textit{rootish}(x) \wedge \textit{leafytopneigh}(x, y, P_1) \wedge \textit{topneigh}(y, b, P_2) \wedge \textit{pbranch}(b)$ says that $b$ is the base of a single cross, $x$ its tail, $y$ its leafy vertex in the middle, and $P_1$ and $P_2$ the paths connecting $x$, $y$, and $b$, respectively;

- $\varphi_{Cr_1}(H) := \exists x, y, b \subseteq V(H) \exists P_1, P_2 \subseteq H\, \big(\textit{cross}_1(x, y, b, P_1, P_2) \wedge \forall e \in H$
  $(e \in P_1 \vee e \in P_2 \vee x \in e \vee y \in e) \wedge \forall e(x \in e \vee y \in e \to e \in H)\big)$ defines a single cross of $\mathcal{W}$; the formula $\varphi_{Cr_2}(H)$ defining a double cross can be obtained analogously; note that since the leaves of $T$ are marked with leaf-marks and $X$ lies flat in $T$, the bases of the crosses are not leafy – this is crucial for making the crosses definable;

- the formula $\varphi_X(v)$ defining the bases of the crosses is immediately derived using $\textit{cross}_1$ and its analog $\textit{cross}_2$;

- $\textit{spec}(v) := \textit{deg}^{=4}(v) \vee \textit{deg}^{=3}(v) \wedge \neg \exists w, P\, \big(\textit{topneigh}(v, w, P) \wedge \textit{leafy}(w)\big)$ defines the set of *special* vertices of $G$, i.e. the vertices of degree 4 or 3 that do not have a leafy vertex as a topological neighbour;

- $\varphi_A(v) := \textit{deg}^{=3}(v) \wedge \neg\textit{spec}(v) \wedge \exists w, P\, \big(\textit{topneigh}(v, w, P) \wedge \textit{spec}(w)\big)$ defines the good vertices of the tree as given in Definition 4.16;

- $\varphi_{PQ}(H) := \forall v, e\big(\varphi_A(v) \wedge v \in e \to (\exists u, P(\textit{topneigh}(v, u, P) \wedge \textit{spec}(u) \wedge e \in P) \leftrightarrow e \in H)\big) \wedge$
  $\forall v, e\big(\neg\varphi_A(v) \wedge v \in e \wedge v \in V(H) \to e \in H\big) \wedge \textit{conn}(H)$ defines the edges of the grid-like minor $(\mathcal{P}, \mathcal{Q})$ by specifying that of the edges adjacent to a good vertex exactly the one that starts a good path belongs to $(\mathcal{P}, \mathcal{Q})$; the definition is completed by taking a maximal connected subgraph that includes these edges;

- $\textit{leafmark}(H) := \exists e_1, e_2, x, y, z\, \big(H = \{e_1, e_2\} \wedge e_1 = \{x, y\} \wedge e_2 = \{y, z\} \wedge \textit{leaf}(x) \wedge \textit{deg}^{=2}(y)\big)$ defines a leaf-mark according to Definition 4.16;

- $\textit{rootmark}(H) := \exists e_1, e_2, r, x, y\, \big(H = \{e_1, e_2\} \wedge e_1 = \{r, x\} \wedge e_2 = \{r, y\} \wedge \textit{leaf}(x) \wedge \textit{leaf}(y)\big)$ defines the marking of the root;

- $\varphi_T(H) := \forall e\big(e \in H \leftrightarrow \neg \exists H'\big(e \in H' \wedge (\textit{rootmark}(H') \vee \textit{leafmark}(H') \vee \varphi_{Cr_1}(H') \vee \varphi_{Cr_2}(H') \vee \varphi_{PQ}(H'))\big)\big)$ uniquely defines the tree $T$ as the set of edges that do not belong to markings, crosses, or the grid-like minor.

Finally, we obtain $\varphi_\leq(x, y) := \exists T\, (\varphi_T(T) \wedge x \in V(T) \wedge y \in V(T) \wedge \varphi_\leq(x, y, T))$ where $\varphi_\leq(x, y, T)$ denotes the formula obtained from Lemma 5.1. □

## 6 MSO$_2$ Interpretations and Walls

In this section, we first show the intractability of MSO$_2$ on walls and then lift this to show the general result. For this, we first recall the well-known fact that MSO$_2$ is intractable on coloured walls. In Section 4, we showed that given a word $w$ and a graph $G$ of large enough treewidth, we can construct either a wall encoding $w$ or a labelled tree-ordered web encoding $w$. For either outcome we will define an MSO$_2$-interpretation of coloured walls in these structures which will allow us to transfer the intractability results from coloured walls to these structures.

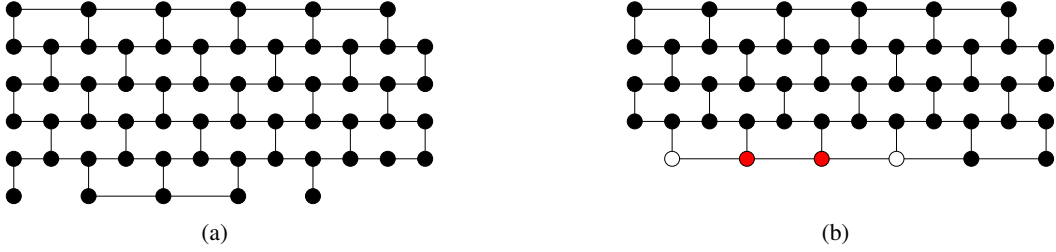

**Figure 8. (a) A wall-encoding of** $0110$ **and (b) the coloured wall it is interpreted as.**

### 6.1 MSO$_2$-Interpretations

We first recall briefly the concepts of interpretations (see e.g. [Hod97]). In logic, they play a similar role to many-one reductions in complexity theory.

**Definition 6.1.** *Let $\sigma$ and $\tau$ be signatures and let $\overline{X}$ be a tuple of monadic second-order variables. An interpretation of $\tau$ in $\sigma$ with parameters $\overline{X}$ is a tuple $\Theta := \bigl(\varphi_{valid}, \varphi_{univ}(x), \varphi_\sim(x, y), (\varphi_R(\overline{x}))_{R \in \tau}\bigr)$ of MSO$_2[\sigma \dot\cup \overline{X}]$-formulas, where the arity of $\overline{x}$ in $\varphi_R(\overline{x})$ is $ar(R)$, such that for all $\sigma$-structures $A$ and assignments $\overline{Y} \subseteq U(A)$ to $\overline{X}$ with $(A, \overline{Y}) \models \varphi_{valid}$, $\varphi_\sim$ defines an equivalence relation on $\varphi_{univ}(A)$.*

For an interpretation $\Theta$ we will denote $\varphi_{valid}$ by $\varphi_{valid}(\Theta)$. With any interpretation $\Theta$ we associate a map taking a $\sigma$-structure $A$ and $\overline{Y} \subseteq U(A)$ such that $(A, \overline{Y}) \models \varphi_{valid}$ to a $\tau$-structure $H$ with universe $U(H) := \varphi_{univ}(A, \overline{Y})_{|\varphi_\sim(A, \overline{Y})} := \{[v]_\sim : (A, \overline{Y}) \models \varphi_{univ}(v)\}$ where $[v]_\sim$ denotes the equivalence class of $v$ under $\varphi_\sim(A, \overline{Y})$. For $R \in \tau$ of arity $r := ar(R)$ we define $R(H) := \{([a_1], \ldots, [a_r]) : (A, \overline{Y}) \models \varphi_R(a_1, \ldots, a_r)\}$. For given $(A, \overline{Y})$, we denote the resulting $\tau$-structure by $\Theta(A, \overline{Y})$.

Furthermore, any interpretation $\Theta$ also defines a translation of MSO$_2[\tau]$-formulas $\varphi$ to MSO$_2[\sigma]$-formulas $\Theta(\varphi)$ by replacing occurrences of relations $R \in \tau$ by their defining formulas $\varphi_R \in \Theta$ in the usual way (see [Hod97] for details) so that the following lemma holds. From now on we will always let $\sigma$ and $\tau$ be $\sigma_{graph}$ or expansions thereof and therefore speak about interpretations without any reference to specific signatures.

**Lemma 6.2** (Interpretation Lemma)**.** *Let $\Theta$ be an MSO$_2$-interpretation with parameters $\overline{X}$. For any $\sigma_{graph}$-structure $A$ and assignment $\overline{Y} \subseteq U(A)$ to $\overline{X}$ s.t. $(A, \overline{Y}) \models \varphi_{valid}(\Theta)$, and any MSO$_2$-sentence $\varphi$ we have $\Theta(A, \overline{Y}) \models \varphi$ if, and only if, $(A, \overline{Y}) \models \Theta(\varphi)$.*

### 6.2 MSO$_2$ on Coloured Elementary Walls

The signature $\sigma_{wall}$ of coloured walls is defined as $\sigma_{wall} := \{V, E, \in, C_0, C_1\}$, where $V, E, C_0, C_1$ are unary relation symbols and $\in$ is a binary relation symbol. A $\sigma_{wall}$-structure $W$ is a *coloured elementary $\ell \times \ell$-wall* if its $\sigma_{graph}$-reduct $W_{|\{V, E, \in\}}$ is an elementary $\ell \times \ell$-wall according to Definition 2.3. $W$ encodes a word $w := w_1 \ldots w_n \in \Sigma^n$ with power $d$ if $\ell > n^d$ and if $\{v_{1,i} : 1 \leq i \leq \ell\}$ are the vertices on the bottom row then $v_{1,i} \in C_0$ if and only if $w_i = 0$ and $v_{1,i} \in C_1$ if and only if $w_i = 1$, for all $1 \leq i \leq n$, and $C_0 \cup C_1 = \{v_{1,i} : 1 \leq i \leq n\}$ (see Figure 8 (b)).

The following lemma, whose proof is standard, is part of the folklore and immediately implies Theorem 6.4 below.

**Lemma 6.3.** *Let $M$ be a nondeterministic $n^d$-time bounded Turing machine. There is a formula $\varphi_M \in$ MSO$_2$ such that for all words $w \in \Sigma^\star$, if $W$ is a coloured elementary wall encoding $w$ with power $d$, then $W \models \varphi_M$ if, and only if, $M$ accepts $w$. Furthermore, the formula $\varphi_M$ can be constructed effectively from $M$. The same holds if $M$ is an alternating Turing machine with a bounded number of alternations, as they are used to define the polynomial-time hierarchy.*

**Theorem 6.4.** *For $d \geq 2$ let $\mathfrak{W}_d$ be the class of coloured elementary walls encoding words with power $d$. Then $\mathrm{MC}(\mathrm{MSO}_2, \mathfrak{W}_d)$ is not in $\mathsf{XP}$ unless $\mathsf{P} = \mathsf{NP}$.*

## 6.3 MSO$_2$ on Uncoloured Walls

The previous paragraph stated the intractability of MSO$_2$ on coloured elementary walls. As one possible outcome of Theorem 4.22 we get an uncoloured wall $W$, not necessarily elementary, of sufficient size. In the absence of colours we will encode a word $w$ in $W$ by taking a suitable subgraph $W_{enc} \subseteq W$ as follows.

Let $w := w_1, \ldots, w_n \in \{0,1\}^*$ be a word of length $n$, let $d \geq 1$ and let $m := n^d + 1$. The aim is to encode $w$ in a wall $W$ of order at least $m \times m$. Let $v_1, \ldots, v_{m+1}$ be the nails (see 2.3) on the bottom row $B \subseteq W$ of $W$ and, for $1 \leq i \leq m$, let $P_i \subseteq B$ be the subpath connecting $v_i$ and $v_{i+1}$. Let $W_{enc} \subseteq W$ be the subgraph obtained from $W$ by deleting the vertices $v_{n+2}, \ldots, v_{m+1}$ and the internal vertices and edges of $P_i$ for each $1 \leq i \leq n$ with $w_i = 0$. All other paths remain unchanged. We say that $W'$ is a *wall-encoding of $w$ with power $d$*. Figure 8 (a) shows a wall-encoding of the word 0100 with power 1. Note that by deleting the vertices $v_{n+2}, \ldots, v_{m+1}$ we ensure that the left side of $W_{enc}$ is uniquely identified.

It is well-known, see e.g. [Kre09b], that a wall $W$ can be defined by an MSO$_2$-formula $\varphi_{wall}(W, \mathcal{R}, \mathcal{C}, B, L, T, R, N)$ expressing that

(i) $\mathcal{R}$ and $\mathcal{C}$ are sets of edges consisting of disjoint paths such that $W = \mathcal{R} \cup \mathcal{C}$; we think of $\mathcal{R}$ as the set of rows and of $\mathcal{C}$ as the set of columns of $W$; note that an $m \times n$-wall has exactly $m + 1$ rows and $n + 1$ columns and each column and row have exactly two nails in common, except on the top and bottom rows, where they intersect in only one nail;

(ii) $B, T \subseteq \mathcal{R}$ are the bottom and top row of $W$ and $L, R \subseteq \mathcal{C}$ are the leftmost and rightmost column of $W$, respectively; and

(iii) $N \subseteq V(W)$ is the set of nails of $W$.

Sometimes we use $\varphi_{wall}(W) := \exists \mathcal{R}, \mathcal{C}, B, L, T, R, N \, \varphi_{wall}(W, \mathcal{R}, \mathcal{C}, B, L, T, R, N)$ as a shortcut. Furthermore, we need the formula $dpaths(H) := ac(H) \wedge \forall v \in V(H) \, deg^{\leq 2}(v)$, which expresses that $H$ is a set of edges consisting of a number of disjoint paths, and the formula $pathof(P, u, v, H) := P \subseteq H \wedge pathends(u, v, P) \wedge leaf(u, H) \wedge leaf(v, H)$ expressing that $P$ is a path with endpoints $v$ and $w$ of degree 1 in $H$; in particular, if $H$ is a set of disjoint paths, this formula asserts that $P$ is one of the disjoint paths in $H$; we use the shortcut $pathof(P, H) := \exists u, v \, pathof(P, u, v, H)$. We can now prove

**Theorem 6.5.** *There is an MSO$_2$-interpretation $\Theta$ of $\sigma_{wall}$ in $\sigma_{graph}$ such that if $W_{enc}$ is an uncoloured wall-encoding of order at least 4 of $w \in \Sigma^*$ with power $d$ then $\Theta(W_{enc})$ is a coloured elementary wall encoding $w$ with power $d$.*

*Proof.* We can think of an $m \times m$ wall-encoding $W_{enc}$ of $w = w_1 \ldots w_n$ as an $(m-1) \times m$ wall $W$ augmented by a set of edges $M$ that we call the *marking of the wall*. We require that the marking is attached only to the bottom row and only to nonnail vertices or to the first, i.e. leftmost vertex on the bottom row. We define the interpretation $\Theta$ with parameters $W, B, L, R, N$, and $M$ as follows. To this end, we make use of a formula $\varphi_{bwall}$ expressing that $W$ is a wall in which the left and right columns each contain 3 or more nails of global degree 2:

$$\varphi_{bwall}(W, B, L, R, N) := \exists \mathcal{R}, \mathcal{C}, T \, \varphi_{wall}(W, \mathcal{R}, \mathcal{C}, B, L, T, R, N) \wedge$$
$$\exists^{\geq 3} v \in N \cap V(L) \, deg^{=2}(v) \wedge \exists^{\geq 3} v \in N \cap V(R) \, deg^{=2}(v)$$

If the given graph is a wall-encoding, this makes sure that the wall spans from the left side to the right side of the wall-encoding and furthermore, cannot contain any edge of the marking of the wall, since any wall that contains such edges cannot contain 3 or more nails of degree 2 on its right boundary. Again, we define the shortcut $\varphi_{bwall}(W)$ similar to the case of $\varphi_{wall}(W)$. Now we can define the formula $\varphi_{valid}(\Theta)$:

$$\begin{aligned}\varphi_{valid}(W, B, L, R, N, M) := \; & \varphi_{bwall}(W, B, L, R, N) \wedge \\ & \forall W'(\varphi_{bwall}(W') \to W' \subseteq W) \wedge \\ & \forall e(e \in M \leftrightarrow e \notin W) \wedge \\ & \forall v \in V(M) \cap V(W) \, \bigl(v \in V(B) \wedge (v \in V(L) \vee v \notin N)\bigr)\end{aligned}$$

Let $W_{col} := \Theta(W_{enc})$ be the coloured wall we aim at in the interpretation. We define the set of vertices of $W_{col}$ simply as the set of nails of $W$: $\varphi_V(x) := x \in N$.

The edges of $W_{col}$ are the equivalence classes of subdivided edges of $W$:

$$\begin{aligned}\varphi_E(x) := \; & e \in W \\ \varphi^e_{eq}(x, y) := \; & \exists u, v \in N \, \exists P \subseteq W \, \bigl(x \in P \wedge y \in P \wedge \mathit{pathends}(u, v, P) \wedge \\ & \forall z \in V(P) \, (z = u \vee z = v \vee z \notin N)\bigr)\end{aligned}$$

Now, we obtain the universe, equivalence relation, and incidence relation of $\Theta$ as:

$$\begin{aligned}\varphi_{univ}(x) := \; & \varphi_V(x) \vee \varphi_E(x) \\ \varphi_{eq}(x, y) := \; & \varphi^e_{eq}(x, y) \\ \varphi_\in(v, e) := \; & \varphi_V(v) \wedge \varphi_E(e) \wedge \exists e' \in W \, \bigl(v \in e' \wedge \varphi^e_{eq}(e, e')\bigr)\end{aligned}$$

It remains to define the colours of $W_{col}$. For a nail $v$ on the bottom row of the wall, consider the closest vertices $u$ and $w$ to its left and right that are incident to an edge of the marking, if existent. Now $v$ is to be interpreted as a 0 if and only if $u$ and $w$ belong to *different* connected components of the marking; and $v$ is to be interpreted as a 1 if $u$ and $w$ belong to the *same* connected component of the marking. In formulas, we have

$$\begin{aligned}\mathit{markingof}(x, X, Y) := \; & x \in N \cap V(B) \, \exists u, v \, \exists P \subseteq B \\ & \bigl(\mathit{pathends}(u, v, P) \wedge u \neq v \wedge x \in V(P) \\ & \mathit{components}(X, M) \wedge \mathit{components}(Y, M) \wedge u \in V(X) \wedge v \in V(Y) \wedge \\ & \forall z \in V(P) \, (z = v \vee z = u \vee \neg \exists e \in M \, z \in e)\bigr) \\ \varphi_{C_0}(x) := \; & \exists X, Y \, (\mathit{markingof}(x, X, Y) \wedge X \neq Y) \\ \varphi_{C_1}(x) := \; & \exists X, Y \, (\mathit{markingof}(x, X, Y) \wedge X = Y)\end{aligned}$$

This finishes the definition of the interpretation $\Theta$. □

### 6.4 MSO$_2$ on Labelled Tree-Ordered Webs

The aim of this section is to show that we can define a coloured elementary wall encoding a word $w$ in a labelled tree-ordered web encoding $w$. The proof follows the basic ideas of a related proof by Kreutzer [Kre09b].

**Theorem 6.6.** *There is an* MSO$_2$-*interpretation $\Theta$ such that if $(G, T, r, A, \mathcal{P}, \mathcal{Q}, X, C)$ is a labelled tree-ordered web encoding a word $w$ with power $d$, then $\Theta(G)$ is a coloured elementary wall encoding $w$ with power $d$.*

*Proof.* We will define the interpretation in a sequence of steps and will illustrate the formulas by a labelled tree-ordered web $\mathcal{W} = (G, T, r, A, \mathcal{P}, \mathcal{Q}, X, C)$ encoding a word $w$ of length $n$ with power $d$. The actual formulas will not depend on $\mathcal{W}$ in any form.

By Lemma 5.2, there exist MSO$_2$-formulas $\varphi_T(X), \varphi_R(x), \varphi_A(x), \varphi_X(x), \varphi_{Cr_1}(X), \varphi_{Cr_2}(X), \varphi_{PQ}(X)$, and $\varphi_\preceq(x,y)$ defining $T$, $r$, $A$, $X$, the single and double crosses of $C$, the edges of the grid-like minor $(\mathcal{P}, \mathcal{Q})$, and the canonical order $\leq_T$, respectively. Essentially, we now have formulas which, on $G$ as above, define the labelled tree-ordered web $\mathcal{W}$. As a shortcut, we define $\varphi_T^e(x) := \exists T \varphi_T(T) \land x \in T$ and $\varphi_T^v(x) := \exists T \varphi_T(T) \land v \in V(T)$ and do similarly for $\varphi_{Cr_1}$, $\varphi_{Cr_2}$, and $\varphi_{PQ}$.

What is left to do is to define formulas which generate a wall from the grid-like minor $(\mathcal{P}, \mathcal{Q})$ so that the bottom row of the wall is connected to the vertices in $A = \{v_1, \ldots, v_n\}$ in the correct order, where $n$ is the length of the word $w$. Our interpretation $\Theta$ is defined with main parameters $\mathcal{P}, \mathcal{Q}$, and $H$ that are intended to be the disjoint paths of the grid-like minor $(\mathcal{P}, \mathcal{Q})$ and a wall in the intersection graph $\mathcal{I}(\mathcal{P}, \mathcal{Q})$, respectively; furthermore, we require parameters $B, L, N$ defining the bottom row, leftmost row, and the nails of $H$, respectively. Note that here we regard $\mathcal{P}$ and $\mathcal{Q}$ as sets of edges comprising disjoint paths each.

We start by defining $\mathcal{I} := \mathcal{I}(\mathcal{P}, \mathcal{Q})$ as follows. The *vertices* of $\mathcal{I}$ are equivalence classes of *edges* of $G$ in the grid-like minor that appear in exactly one of $\mathcal{P}$ or $\mathcal{Q}$ and are equivalent if they belong to the same path in $\mathcal{P}$ or $\mathcal{Q}$. The *edges* of $\mathcal{I}$ are equivalence classes of *vertices* of $G$ where two vertices are equivalent if they belong to the intersection of the same pair $P \in \mathcal{P}$ and $Q \in \mathcal{Q}$. Formally, let *pathofPQ*$(P) :=$ *pathof*$(P, \mathcal{P}) \lor$ *pathof*$(P, \mathcal{Q})$ and

$$\varphi_V^\mathcal{I}(x) := (x \in \mathcal{P} \land x \notin \mathcal{Q}) \lor (x \in \mathcal{Q} \land x \notin \mathcal{P})$$
$$\varphi_E^\mathcal{I}(x) := x \in V(\mathcal{P}) \cap V(\mathcal{Q})$$
$$\varphi_{eq}^{\mathcal{I},V}(x,y) := \exists P \big(\textit{pathofPQ}(P) \land \{x,y\} \subseteq P\big)$$
$$\varphi_{eq}^{\mathcal{I},E}(x,y) := \exists P, Q \big(\textit{pathofPQ}(P) \land \textit{pathofPQ}(Q) \land \{x,y\} \subseteq V(P) \cap V(Q)\big)$$
$$\varphi_{eq}^\mathcal{I}(x,y) := \varphi_{eq}^{\mathcal{I},V}(x,y) \lor \varphi_{eq}^{\mathcal{I},E}(x,y)$$
$$\varphi_\in^\mathcal{I}(x,y) := \varphi_V^\mathcal{I}(x) \land \varphi_E^\mathcal{I}(y) \land \exists x', y' \big(\varphi_{eq}(x,x') \land \varphi_{eq}(y,y') \land y' \in x'\big)$$

We would like to express that $H$ is a wall in $\mathcal{I}(\mathcal{P}, \mathcal{Q})$. As described in the previous subsection, we have a formula $\varphi_{wall}$ that asserts a graph to be a wall; if in this formula, we replace every occurrence of $x \in V$ by $\varphi_V^\mathcal{I}(x)$, every occurrence of $x \in E$ by $\varphi_E^\mathcal{I}(x)$, and every occurrence of $v \in e$ by $\varphi_\in^\mathcal{I}(v,e)$, we can derive a formula $\varphi_{wall}^\mathcal{I}(H, B, L, N)$ that asserts that $H$ is a wall in $\mathcal{I}$ with bottom row $B$, leftmost row $B$, and nails $N$; the formula can be slightly adapted in such a way that for each equivalence class that is to be included in one of these sets, all representatives are present. Note that $B$ and $L$ are sets of vertices of $G$, and $N$ is a set of edges of $G$.

Similarly, let $\varphi_{pathends}^\mathcal{I}(x,y,P)$ be a formula derived from the formula *pathends* defined earlier expressing that $P$ is a path in $\mathcal{I}$ with endpoints $x$ and $y$; here, $x$ and $y$ are edges of $G$ is $P$ is a set of vertices of $G$. For notational convenience, let us also write $x \in^\mathcal{I} V(Y)$ for the formula $\exists y \in Y \, \varphi_\in^\mathcal{I}(x,y)$ if $x$ is a vertex of $\mathcal{I}$ and $Y$ is a set of edges of $\mathcal{I}$. Now, we can define that a nail in $B$ is left of another nail in $B$ as follows;

$$\varphi_{left}^\mathcal{I}(x,y) := \exists z \, \exists P \subseteq B(z \in B \cap L \land \varphi_{pathends}^\mathcal{I}(z,y,P) \land x \in^\mathcal{I} V(P))$$

Next, we would like to define that a nail of $H$ is attached to a good vertex $v \in A$:

$$\varphi_{attached}(x,v) := x \in N \land \varphi_A(v) \land \exists u, P(\textit{pathof}(P,v,u,\mathcal{P}) \land x \in P)$$

Now we are ready to define $\varphi_{valid}(\Theta)$:

$$\varphi_{valid}(\mathcal{P}, \mathcal{Q}, H, B, L, N) := dpaths(\mathcal{P}) \wedge dpaths(\mathcal{Q}) \wedge \exists Z \, (\varphi_{PQ}(Z) \wedge \mathcal{P} \cup \mathcal{Q} \subseteq Z) \wedge$$
$$\varphi_{wall}^{\mathcal{I}}(H, B, L, N) \wedge \forall v \, (\varphi_A(v) \to \exists x \, (x \in N \wedge x \in V(B) \wedge \varphi_{attached}(x, v))) \wedge$$
$$\forall x, y \in N \, (x \in V(B) \wedge \exists v \, \varphi_{attached}(x, v) \wedge \varphi_{left}^{\mathcal{I}}(y, x) \to \exists w \, \varphi_{attached}(y, w)) \wedge$$
$$\forall x, y \in V(B) \, \forall u, v \, (\varphi_{attached}(x, u) \wedge \varphi_{attached}(y, v) \to (\varphi_{left}^{\mathcal{I}}(x, y) \leftrightarrow \varphi_{\leq}(u, v)))$$

The first line says that $\mathcal{P}$ and $\mathcal{Q}$ are each sets of edges consisting of disjoint paths and that they are included in the grid-like minor of $G$; the second line asserts that $H$ is a wall in the intersection graph $\mathcal{I}(\mathcal{P}, \mathcal{Q})$ and that all good vertices of the labelled tree-ordered web are attached to some nails of the first row of this wall; the third line ensures that for any nail that is attached to the tree, all the nails to its left are also attached; since there are exactly $n$ good vertices and different nails can only be attached to different good vertices, this implies that exactly the first $n$ nails on the bottom row of $H$ are attached to the tree; finally, the last line asserts that attachments respect the order of the tree.

Now we can easily define the following parts of the interpretation $\Theta$:

$$\varphi_V(x) := x \in N$$
$$\varphi_E(x) := x \in H$$
$$\varphi_{univ}(x) := \varphi_V(x) \vee \varphi_E(x)$$
$$\varphi_{eq}(x, y) := \varphi_{eq}^{\mathcal{I}}(x, y) \vee \exists u, v \in N \, \exists P \subseteq H \, (\varphi_{pathends}^{\mathcal{I}}(u, v, P) \wedge$$
$$\forall z \, (z \in V(P) \to z = u \vee z = v \vee z \notin N) \wedge x \in V(P) \wedge y \in V(P))$$
$$\varphi_{\in}(x, y) := \varphi_V(x) \wedge \varphi_E(y) \wedge \exists x', y' \, (\varphi_{eq}(x, x') \wedge \varphi_{eq}(y, y') \wedge y' \in x')$$

Finally, it remains to define the colours, which we do as follows:

$$\varphi_{crossbase}(x, b) := x \in N \wedge x \in V(B) \wedge \varphi_X(b) \wedge \exists v \, (\varphi_{attached}(x, v) \wedge$$
$$\varphi_{\leq}(b, v) \wedge \forall u \, (\varphi_A(u) \wedge \varphi_{\leq}(u, v) \to \varphi_{\leq}(u, b)))$$
$$\varphi_{C_0}(x) := \exists b \, \exists C \, (\varphi_{crossbase}(x, b) \wedge \varphi_{Cr_1}(C) \wedge b \in V(C))$$
$$\varphi_{C_1}(x) := \exists b \, \exists C \, (\varphi_{crossbase}(x, b) \wedge \varphi_{Cr_2}(C) \wedge b \in V(C))$$

The formula $\varphi_{crossbase}$ determines if the given element $x$ is a nail on the bottom row of the interpreted wall and is attached to some good vertex $v$; in this case, it ensures that $b$ is the base of the cross immediately to the left of $v$ in the tree. Then the formulas $\varphi_{C_0}$ and $\varphi_{C_1}$ simply check if this base $b$ is incident to a single or a double cross. □

## 7 Proof of the Main Theorem

In this section we complete the proof of Theorem 1.2. We first prove Part 1.

Suppose, $MC(MSO_2, \mathcal{C}) \in \mathsf{XP}$, i.e. there is a computable function $f : \mathbb{N} \to \mathbb{N}$ such that given $G \in \mathcal{C}$ and $\varphi \in MSO_2$ we can decide $G \models \varphi$ in time $\mathcal{O}(|G|^{f(|\varphi|)})$. Let $M$ be a nondeterministic Turing machine deciding SAT in quadratic time and let $\varphi_M$ be the formula constructible from $M$ as defined in Lemma 6.3.

Let $\Theta_1$ be the interpretation from the Theorem 6.5 and let $\Theta_2$ be the interpretation from Theorem 6.6. Define $\varphi_M^1 := \Theta_1(\varphi_M)$ and $\varphi_M^2 := \Theta_2(\varphi_M)$.

Let $w \in \{0, 1\}^*$ be a word of which we want to decide whether $w \in$ SAT, let $\ell := |w|$ and $t := 2c\ell^{28}$, where $c$ is the constant from Theorem 4.22. As the treewidth of $\mathcal{C}$ is strongly unbounded by $\log^{28\gamma} n$, there are $\varepsilon < 1$ and a polynomial $p(n)$ of degree less than $\gamma$ such that $\mathcal{C}$ contains a graph $G$ with $\mathrm{tw}(G) \geq \log^{28\gamma} |G|$ and $t \leq \mathrm{tw}(G) \leq p(t)$ and $G$ can be computed in time $2^{|w|^\varepsilon}$; note that this implies that

$$|G| \leq 2^{p(2c\ell 28)^{\frac{1}{28\gamma}}} \leq 2^{|w|^\delta}, \text{ for some } \delta < 1\,.$$

By Theorem 4.22, $G$ either contains a) a wall $W_w$ encoding $w$ with power 2 as a subgraph or b) a labelled tree-ordered web $\mathcal{W} = (H, T, r, A, \mathcal{P}, \mathcal{Q}, X, C)$ encoding $w$ with power 2. Note that we need an encoding with power 2 because $M$ needs $|w|^2$ space cells and computation steps to decide $w \in \mathrm{SAT}$.

In case $a)$, as $\mathcal{C}$ is closed under subgraphs, $W_w \in \mathcal{C}$ and we can therefore decide $W_w \models \varphi_M^1$ in time

$$|W_w|^{f(|\varphi_M^1|)} \leq |G|^{f(|\varphi_M^1|)} \leq (2^{|w|^\delta})^{f(|\varphi_M^1|)} = 2^{f(|\varphi_M^1|)|w|^\delta} = 2^{o(|w|)}\,.$$

By construction, $W_w \models \varphi_M$ if, and only if, $M$ accepts $w$ if, and only if, $w \in \mathrm{SAT}$.

In case $b)$, $H \in \mathcal{C}$ as $H$ is a subgraph of $G$. We can therefore decide $H \models \varphi_M^2$ in time

$$|H|^{f(|\varphi_M^2|)} \leq |G|^{f(|\varphi_M^2|)} \leq (2^{|w|^\delta})^{f(|\varphi_M^2|)} = 2^{f(|\varphi_M^2|)|w|^\delta} = 2^{o(|w|)}$$

By construction, $H \models \varphi_M$ if, and only if, $M$ accepts $w$ if, and only if, $w \in \mathrm{SAT}$.

Hence, in both cases we can decide $w \in \mathrm{SAT}$ in time $2^{o(|w|)}$. This shows Part 1.

To show Part 2, we use the same proof idea. Let $P$ be a language in the polynomial-time hierarchy and let $M$ be an alternating Turing machine with bounded alternation deciding $P$ in time $n^k$. We use essentially the same proof as above but, given a word $w$, we construct a graph $G$ which contains a wall or a labelled tree-ordered web encoding $w$ with power $k$. The rest follows then as before.

This concludes the proof of Theorem 1.2. It is easily seen that the proof can be adapted to classes of graphs closed under spanning subgraphs, i.e. edge deletion: instead of taking subgraphs we simply delete all edges no longer needed and make the $\mathrm{MSO}_2$-formulas ignore isolated vertices.

**Corollary 7.1.** *Let $\mathcal{C}$ be a class of graphs closed under spanning subgraphs, i.e. $G \in \mathcal{C}$, $V(H) = V(G)$, and $E(H) \subseteq E(G)$ implies $H \in \mathcal{C}$.*

1. *If the treewidth of $\mathcal{C}$ is strongly unbounded by $\log^{28\gamma} n$, where $\gamma > 1$ is larger than the gap-degree of $\mathcal{C}$, then $\mathrm{MC}(\mathrm{MSO}_2, \mathcal{C})$ is not in $\mathsf{XP}$, and hence not fixed-parameter tractable, unless $\mathrm{SAT}$ can be solved in subexponential time $2^{o(n)}$.*

2. *If the treewidth of $\mathcal{C}$ is strongly unbounded polylogarithmically then $\mathrm{MC}(\mathrm{MSO}_2, \mathcal{C})$ is not in $\mathsf{XP}$ unless all problems in the polynomial-time hierarchy can be solved in subexponential time.*

## 8. Conclusion

We have presented a strong intractability result for $\mathrm{MSO}_2$ on graph classes of unbounded treewidth. In comparison to Courcelle's theorem, Courcelle's theorem requires the treewidth to be constant whereas our result refers to classes whose treewidth is essentially not bounded logarithmically. As the examples in [MM03] show, there are classes of graphs of unbounded treewidth, closed under subgraphs, which admit tractable $\mathrm{MSO}_2$-model-checking. On the other hand, this is very unlikely to be the case for all classes of logarithmic treewidth. Exploring tractability and intractability of $\mathrm{MSO}_2$ on classes of unbounded treewidth, but bounded by $\log n$, might lead to interesting new results on the boundary of $\mathrm{MSO}_2$-tractability.

The results reported in this part of the thesis refer to $\mathrm{MSO}_2$, i.e. MSO with quantification over sets of edges. For MSO without edge set quantification, referred to as $\mathrm{MSO}_1$, it can be shown that $\mathrm{MSO}_1$ is tractable on any class $\mathcal{C}$ of graphs of bounded *cliquewidth*. Again, except for the examples in [MM03], not much is known about $\mathrm{MSO}_1$ and graph classes of unbounded cliquewidth and it would be very interesting to establish similar results as in this work for the case of cliquewidth. This, however, is much more difficult as no good obstruction similar to grid-like minors is known for this case.